\documentclass[reprint, aps, prd, letterpaper, noshowpacs, amsmath, %
amssymb, amsfonts, nofootinbib, floatfix, twoside, colorlinks]{revtex4-1}
\pdfoutput=1

\usepackage{hyperref}

\usepackage{mathrsfs} 
\usepackage{bm} 
\usepackage[varg]{txfonts} 
\usepackage{tgtermes} 
\usepackage{microtype} 
\makeatletter
\def\MT@register@subst@font{\MT@exp@one@n\MT@in@clist\font@name\MT@font@list
  \ifMT@inlist@\else\xdef\MT@font@list{\MT@font@list\font@name,}\fi}
\makeatother

\renewcommand{\small}{\fontsize{9.45}{10.395}\selectfont}

\renewcommand{\tiny}{\fontsize{5}{6}\selectfont}
\DeclareMathSizes{10.5}{11.0}{7.7}{5.5}
\DeclareMathSizes{9.45}{9.5}{6.7}{4.8}
\DeclareMathSizes{8.4}{8.5}{6.0}{4.3}

\usepackage{graphicx} %
\usepackage{color} %
\usepackage{xspace}
\usepackage{braket}
\usepackage{accents}
\usepackage{siunitx}

\definecolor{CiteColor}{rgb}{0.18039, 0.18824, 0.57255}
\definecolor{UrlColor} {rgb}{0.741, 0.173, 0.000}
\definecolor{LinkColor}{rgb}{0.25098, 0.47843, 0.04706}

\makeatletter

\def\@seccntformat#1{\csname the#1\endcsname.~}%

\def\section{%
  \@startsection {section}
  {1} {\z@} {0.55cm \@plus1ex \@minus .02ex}%
    {0.225cm} { \normalfont\bfseries \centering}%
}%
\def\subsection{%
  \@startsection {subsection}
  {2} {\z@ } {0.45cm \@plus 0.8ex \@minus 0.2ex}%
  {0.1125cm}{\normalfont \bfseries \centering }}
\def\subsubsection{%
  \@startsection {subsubsection}
  {3} {\z@ } {0.4cm \@plus 0.6ex \@minus 0.1ex}%
  {0.075cm}{\normalfont \it \centering }}

\newcommand{\surnames}[1]{\def\@surnamelist{#1}\relax}
\ifx \@shorttitle \@empty
\else
  \def\@oddhead{\small \MakeUppercase{\@shorttitle} \hfill}
\fi
\ifx \@surnamelist \@empty
\else
  \def\@evenhead{\small \MakeUppercase{\@surnamelist} \hfill}
\fi
\def\@oddfoot{\reset@font\hfil\thepage\hfil}%
\def\@evenfoot{\reset@font\hfil\thepage\hfil}%
\g@addto@macro\maketitle{\global\@specialpagetrue\gdef\@specialstyle{plain}}

\makeatother

\renewcommand{\d}{\ensuremath{\mathrm{d}}}
\newcommand{\e}{\ensuremath{\mathrm{e}}}
\let\Originalidefinition\i
\renewcommand{\i}{\ensuremath{\mathrm{i}}}

\renewcommand{\c}{\mathrm{c}}
\newcommand{\G}{\mathrm{G}}
\newcommand{\MSun}{\ensuremath{M_\odot}\xspace}

\newcommand{\Mtot}{\ensuremath{M_\text{tot}}}
\DeclareSIUnit{\strain}{strain}
\DeclareSIPrePower\root{1/2}
\DeclareSIUnit{\parsec}{pc}
\DeclareSIUnit{\SolarMass}{\ensuremath{\MSun}}
\DeclareSIUnit{\Mass}{\ensuremath{M}}

\newcommand{\abs} [1]{\left\lvert{#1}\right\rvert}

\DeclareMathOperator{\sgn}{sgn}

\newcommand{\define}{\coloneqq}

\newcommand{\R}{\ensuremath{\mathcal{R}} }

\newcommand{\mTwoYlm}[1]{\ensuremath{\scripts{_{-2}}{Y}{_{#1}}}}

\newcommand{\htilde}{\ensuremath{{\tilde{h}}}}

\newcommand{\omegah}{\ensuremath{\omega_{\text{hyb}}}}
\newcommand{\fhyb}{\ensuremath{f_{\text{hyb}}}}
\newcommand{\thyb}{\ensuremath{t_{\text{hyb}}}}

\newcommand{\Pade}{{Pad{\'{e}}}\xspace}

\definecolor{NoteColor}{rgb}{0.900, 0.218, 0.000}

\definecolor{NewColor}{rgb}{0,.55,0}

\makeatletter
\newcommand{\CapName}[1]{\textbf{#1}.}
\newcommand{\ShowDimensions}{%
  \typeout{The font encoding is \f@encoding}        %
  \typeout{The font family is \f@family}            %
  \typeout{The font series is \f@series}            %
  \typeout{The font shape is \f@shape}              %
  \typeout{The font size is \f@size}                %
  \typeout{The baselineskip is \f@baselineskip}     %
  \typeout{The math font size is \tf@size}          %
  \typeout{The math script size is \sf@size}        %
  \typeout{The math scriptscript size is \ssf@size} %
  \typeout{The linewidth is \the\linewidth}         %
}
\newcommand{\prefixscripts}[2]{%
  \@mathmeasure\z@\displaystyle{#2}%
  \global\setbox\@ne\vbox to\ht\z@{}\dp\@ne\dp\z@
  \setbox\tw@\box\@ne
  \@mathmeasure4\displaystyle{\copy\tw@#1}%
  \@mathmeasure6\displaystyle{#2}%
  \dimen@-\wd6 \advance\dimen@\wd4 \advance\dimen@\wd\z@
  \hbox to\dimen@{}{\kern-\dimen@\box4\box6}%
}
\newcommand{\scripts}[3]{%
  \@mathmeasure\z@\displaystyle{#2}%
  \global\setbox\@ne\vbox to\ht\z@{}\dp\@ne\dp\z@
  \setbox\tw@\box\@ne
  \@mathmeasure4\displaystyle{\copy\tw@#1}%
  \@mathmeasure6\displaystyle{#2#3}%
  \dimen@-\wd6 \advance\dimen@\wd4 \advance\dimen@\wd\z@
  \hbox to\dimen@{}{\kern-\dimen@\box4\box6}%
}
\makeatother

\makeatletter
\let\protect\relax
{\catcode`\|=\active
  \xdef\InnerProduct{\protect\expandafter\noexpand\csname InnerProduct \endcsname}
  \expandafter\gdef\csname InnerProduct \endcsname#1{%
    \begingroup
    \ifx\SavedDoubleVert\relax
    \let\SavedDoubleVert\|\let\|\IpDoubleVert
    \fi
    \mathcode`\|32768\let|\IPVert
    \left({#1}\right)
    \endgroup
  }
}
\def\IPVert{\@ifnextchar|{\|\@gobble}
     {\egroup\,\mid@vertical\,\bgroup}}
\def\IPDoubleVert{\egroup\,\mid@dblvertical\,\bgroup}
\let\SavedDoubleVert\relax
\def\midvert{\egroup\mid\bgroup}
\def\SetVert{\@ifnextchar|{\|\@gobble}
    {\egroup\;\mid@vertical\;\bgroup}}
\def\SetDoubleVert{\egroup\;\mid@dblvertical\;\bgroup}
\def\mid@vertical{\mskip1mu\vrule\mskip1mu}
\def\mid@dblvertical{\mskip1mu\vrule\mskip2.5mu\vrule\mskip1mu}
\makeatother
\newcommand{\Overlap}{\Braket}
\newcommand{\Mismatch}{\ensuremath{\text{MM}}}
\newcommand{\MM}[2]{\ensuremath{\Mismatch\left(#1,#2\right)}}
\newcommand{\TargetMismatch}{\ensuremath{\Mismatch_{\text{target}}}}

\newcommand{\Cornell}{\affiliation{Center for Radiophysics and
    Space Research, Cornell University, Ithaca, New York 14853, USA}}%

\hypersetup{linkcolor=LinkColor}
\hypersetup{citecolor=CiteColor}
\hypersetup{urlcolor=UrlColor}

\begin{document}


\graphicspath{%
  {Plots/}%
  {Plots/MismatchContours/}%
  {Plots/MismatchVersusOmegah/}%
  {Plots/EOBvsPN/}%
}

\title[Uncertainty in hybrid gravitational waveforms] {Uncertainty in
  hybrid gravitational waveforms: \texorpdfstring{\\}{} Optimizing
  initial orbital frequencies for binary black-hole simulations}

\surnames{Boyle}

\author{Michael Boyle} \Cornell

\date{\today}

\begin{abstract}
  A general method is presented for estimating the uncertainty in
  hybrid models of gravitational waveforms from binary black-hole
  systems with arbitrary physical parameters, and thence the highest
  allowable initial orbital frequency for a numerical-relativity
  simulation such that the combined analytical and numerical waveform
  meets some minimum desired accuracy.  The key strength of this
  estimate is that no prior numerical simulation in the relevant
  region of parameter space is needed, which means that these
  techniques can be used to direct future work.  The method is
  demonstrated for a selection of extreme physical parameters.  It is
  shown that optimal initial orbital frequencies depend roughly
  linearly on the mass of the binary, and therefore useful accuracy
  criteria must depend explicitly on the mass.  The results indicate
  that accurate estimation of the parameters of stellar-mass
  black-hole binaries in Advanced LIGO data or calibration of
  waveforms for detection will require much longer numerical
  simulations than are currently available or more accurate
  post-Newtonian approximations---or both---especially for
  comparable-mass systems with high spin.
\end{abstract}



\maketitle


Numerical relativists face a thorny dilemma when creating initial data
for simulations of binary black holes.  Two competing motivations vie
for control of one key parameter: the initial orbital frequency of the
binary, $\Omega_{0}$.  On one hand, the larger the initial frequency
is, the more quickly the simulation will run.  In fact, at lowest
order, doubling $\Omega_{0}$ will shorten a simulation by a factor of
six.  On the other hand, the smaller the initial frequency is, the
clearer the correspondence will be between the numerical simulation
and the system found in nature.  Post-Newtonian approximations will be
more accurate; the velocity and spin of each black hole will be more
well defined and easily measured; even the junk radiation will be
smaller~\cite{LovelaceEtAl:2008}.  In practice, numerical relativists
choose $\Omega_{0}$ largely by intuition, with primary considerations
being the available computer time and the roundness of the number.
This lack of precision can lead to simulations that are too short and
should ideally be redone, or are longer than necessary---a waste of
resources in either case.  More objective choices are possible, and
will be needed to improve the effectiveness of numerical relativity.
This paper demonstrates a technique\footnote{The technique expands on
  one introduced in Ref.~\cite{Boyle:2010}---which was partially
  implemented in Ref.~\cite{DamourEtAl:2011}---to apply to the
  time-domain methods currently in use by most numerical relativists,
  to use the most accurate inspiral models available, and to include
  more general accuracy requirements.  The results obtained here
  broadly agree with the results of Refs.~\cite{SantamariaEtAl:2010,
    HannamEtAl:2010, MacDonaldEtAl:2011}, which test whether completed
  numerical simulations are long enough.}  %
to estimate the optimal value of $\Omega_{0}$, even for systems in
unexplored regions of parameter space.

The field of numerical relativity (NR) exists because of the failure
of post-Newtonian models (PN); at some point NR must take over from
PN.  Of course, PN approximations don't simply break down at one
catastrophic instant, having been perfectly accurate before.  Rather,
the approximations gradually deteriorate as they approach merger.  The
question of exactly where NR needs to replace PN is thus a question of
how accurate the model needs to be.  In the context of designing model
waveforms for detection, we are given precise objectives.  This allows
us to \emph{quantitatively} resolve the conflict between decreasing
the length of a simulation and improving the quality of the final
modeled waveform.  The quality of the final waveform is impacted by
the accuracy of both PN and NR data.  Because we already have the PN
data that will be used in the final waveform, we can test how much of
it can be used if we are to achieve a target accuracy.  Where that
ends is where the NR simulation must begin.

Estimating the impact of PN errors depends on understanding how the
data will be used in the finished product.  In this context, that
means understanding waveforms used in data analysis for
gravitational-wave detectors.  Advanced detectors of the near future
will be more sensitive over broader ranges of frequencies than current
detectors, which tightens the requirements for accurate modeling of
physical waveforms~\cite{FlaminioEtAl:2005, Shoemaker:2010}.  In
particular, analysis of detector data will require waveforms that are
not only more precisely coherent, but also coherent over a greater
range of frequencies.  PN waveforms alone are expected to be
sufficient \emph{for detection} up to a total system mass
somewhere\footnote{The exact mass delimiting the range of validity
  depends, of course, on the parameters of the particular system and
  the detector in question.  Moreover, this says nothing about
  parameter estimation.  The methods of this paper are primarily
  applicable to parameter estimation, but do have some relevance to
  detection.}  %
around \SI{12}{\SolarMass}~\cite{BuonannoEtAl:2009}, and are expected
to be entirely inadequate for parameter estimation.  Exclusively
numerical waveforms, on the other hand, are limited in their
usefulness to high-mass systems; current numerical simulations can
only cover the Advanced LIGO band down to \SI{10}{\hertz} for masses
greater than about \SI{100}{\SolarMass}~\cite{BoyleEtAl:2009}.  This
leaves a large gap in an astrophysically interesting
range~\cite{FarrEtAl:2010}.  To close the gap using NR alone would
require simulations roughly \num{250} times longer than are currently
available---or more for parameter estimation.  Notwithstanding
possible dramatic improvements to the efficiency of long
simulations~\cite{LauEtAl:2009, HennigAnsorg:2009, LauEtAl:2011}, this
will be impractical for some time to come, especially for large
surveys of parameter space~\cite{NRAR:2011}.

\begin{figure}
  \includegraphics[width=\linewidth]{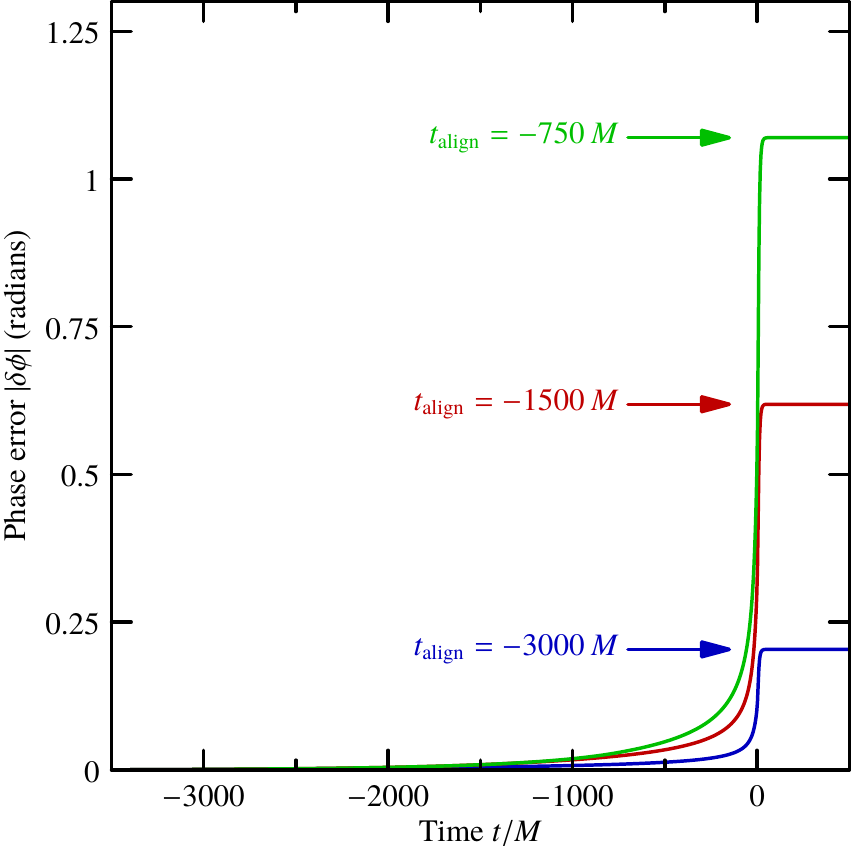}
  \caption{\CapName{Error caused by aligning at late times} This plot
    shows the phase error in hybrid waveforms created by aligning an
    ersatz NR waveform (EOB) to a PN model at various times, relative
    to the hybrid when aligned very far in the past.  As the alignment
    point is moved closer to merger ($t/M=0$), the total phase error
    increases because the hybrid waveform incorporates more and more
    of the inaccurate PN data.  Note that, in each case, the error
    grows most rapidly near merger.  If the merger occurs at a
    frequency to which the detector is sensitive, this phase error
    will negatively impact the match found by data analysts (see
    Sec.~\ref{sec:Motivation}).  Evaluating that impact using
    artificial NR data is the essence of the method presented in this
    paper. \label{fig:HybridMatching}}
\end{figure}

Instead of running such dramatically long simulations, we
``hybridize''---synthesizing a single coherent waveform by combining
the long PN inspiral with the short NR merger and ringdown.  In the
interest of simplicity of presentation, let us assume for the moment
that a hybrid waveform is constructed by aligning the PN and NR
waveforms at a single point in time---using PN data before that point
and NR data after---and that the NR portion is essentially perfect.
Now, as the alignment point moves closer to merger, the hybrid
waveform will become less accurate since it includes more of the
deteriorating PN waveform, as demonstrated in
Fig.~\ref{fig:HybridMatching}.  In fact, because the hybridization
procedure uses only the worst data when aligning at very late times,
errors in the PN waveform just prior to the alignment point will
unavoidably taint the data, meaning that the long inspiral and the
merger will be dephased relative to the physically correct waveform.
The error in the complete hybrid, then, depends crucially on the error
in the PN waveform and particularly on the growth in the error at late
times.

We cannot know the \emph{error} in a PN waveform---being the
difference from the unknown correct waveform it attempts to
model---without unreasonably long numerical simulations.  We can,
however, estimate our \emph{uncertainty} by creating a range of
equally plausible PN waveforms.  We can then attach these to some
ersatz NR waveform,\footnote{As shown later in this paper, the final
  results will not depend strongly on the particular choice of ersatz
  NR waveform.}  %
such as effective one-body (EOB) or phenomenological
waveforms~\cite{BuonannoDamour:1999, BuonannoDamour:2000,
  BuonannoEtAl:2009a, PanEtAl:2010, AjithEtAl:2007, AjithEtAl:2009,
  SturaniEtAl:2010}, to form a range of plausible hybrids.  We can
compare that range quantitatively to the given error budget.  By
repeating this process using each of many possible hybridization
frequencies, we can discover which hybridization frequencies produce
waveforms that satisfy the error budget.  The highest such frequency
minimizes the length of the simulation, and is thus the optimal value.

To summarize, the work to be done in this scheme consists of three
basic steps:
\begin{list}{\labelenumi}{\usecounter{enumi}\leftmargin=1.1em}
 \item Choose an appropriate ersatz NR waveform and construct a range
  of plausible inspiral waveforms
  (Sec.~\ref{sec:CreatingTheWaveforms});
 \item Hybridize the inspiral and ersatz NR waveforms, evaluate the
  mismatches among them, and repeat for various hybridization
  frequencies (Sec.~\ref{sec:EvaluatingUncertainty});
 \item Choose $\Omega_{0}$ to agree with the highest hybridization
  frequency that satisfies the error requirements
  (Sec.~\ref{sec:UsingTheOverlapsToFindOmegaNought}).
\end{list}
This, of course, assumes that there is some freedom in choosing the
initial orbital frequency, as will be the case for NR groups setting
out to run a simulation; there is always a choice between running many
short simulations and fewer long simulations---or requesting a larger
allocation of computer time.  This method is designed to help in
making that choice.  On the other hand, given a completed waveform,
steps \num{1} and \num{2} above can be used to evaluate the
uncertainty in the resulting hybrids.  Finally, understanding the
results can aid in the design of reasonable and effective accuracy
goals, before any simulation is undertaken.

The method will be demonstrated for a few interesting cases, probing
the ``corners'' of a simple parameter space: equal-mass nonspinning,
equal-mass high-spin, large mass-ratio nonspinning, and large
mass-ratio high-spin systems.  The key result of this paper is the
plot of the uncertainty of the hybrids for those systems,
Fig.~\ref{fig:MismatchContours}, discussed in
Sec.~\ref{sec:TheCornersOfParameterSpace}.  Finally,
Sec.~\ref{sec:Conclusions} summarizes the conclusions and outlines
possible applications and extensions to this method.

The general method presented here can be applied quite broadly.
However, to demonstrate the method, we need to make several specific
choices.  Details of the implementation are given in
Appendix~\ref{sec:DetailsOfTheImplementation}.  In particular, this
paper uses a simplified EOB model, described in
Appendix~\ref{sec:EOBModel}, to supply the ersatz NR waveforms.  In
Appendix~\ref{sec:ImportanceOfErsatzNRWaveform}, the results for the
equal-mass nonspinning case are redone using actual numerical data, to
check---to the extent possible---that the final results are not
sensitive to this choice for the ersatz NR waveform.  The uncertainty
will be measured by a range of plausible waveforms formed by
hybridizing the EOB merger and ringdown with the inspiral portion of
EOB and TaylorT1--T4 approximants~\cite{DamourEtAl:2001,
  BuonannoEtAl:2007, BoyleEtAl:2007} using all known information (full
PN orders), as recently recalculated and summarized by members of the
NINJA-2 collaboration~\cite{NINJA2:2011}.  Specifically, the amplitude
includes terms up to 3.0\,PN order, and the phase includes terms up to
3.5\,PN order.  See Appendix~\ref{sec:PNModel} for more detail.  The
hybridization will be done in the time domain by aligning at
particular frequencies~\cite{BakerEtAl:2007, BoyleEtAl:2007}, then
blending the PN and NR waveforms as described in
Ref.~\cite{BoyleEtAl:2009}.  The resulting hybrids will be compared
along the positive $z$ axis, computing the match [see
Eq.~\eqref{eq:Overlap}] using the Advanced LIGO high-power noise curve
with no detuning~\cite{Shoemaker:2010}, scaling the total system mass
between \SIlist{5;50}{\SolarMass}.

Throughout this paper, the uppercase Greek letters $\Phi$ and $\Omega$
refer to the orbit of a binary, contrasting with the lowercase Greek
letters $\phi$ and $\omega$ which refer to the phase and frequency of
the emitted gravitational waves.  Unless otherwise specified, $\phi$
and $\omega$ refer to the $(\ell,m) = (2,2)$ mode in a spin $s=-2$
spherical harmonic decomposition of the gravitational wave.

\section{Creating the waveforms}
\label{sec:CreatingTheWaveforms}
The first task before us is to construct a large group of model
waveforms to be compared to each other.  In this section, a review of
the standard error measure used in data analysis motivates the use of
complete inspiral-merger-ringdown waveforms for our tests, while
finding encouraging signs that high accuracy of these waveforms is not
essential.  We then examine in greater detail the construction of a
credibly broad selection of PN waveforms.

\subsection{Motivation}
\label{sec:Motivation}
\begin{figure}
  \includegraphics[width=\linewidth]{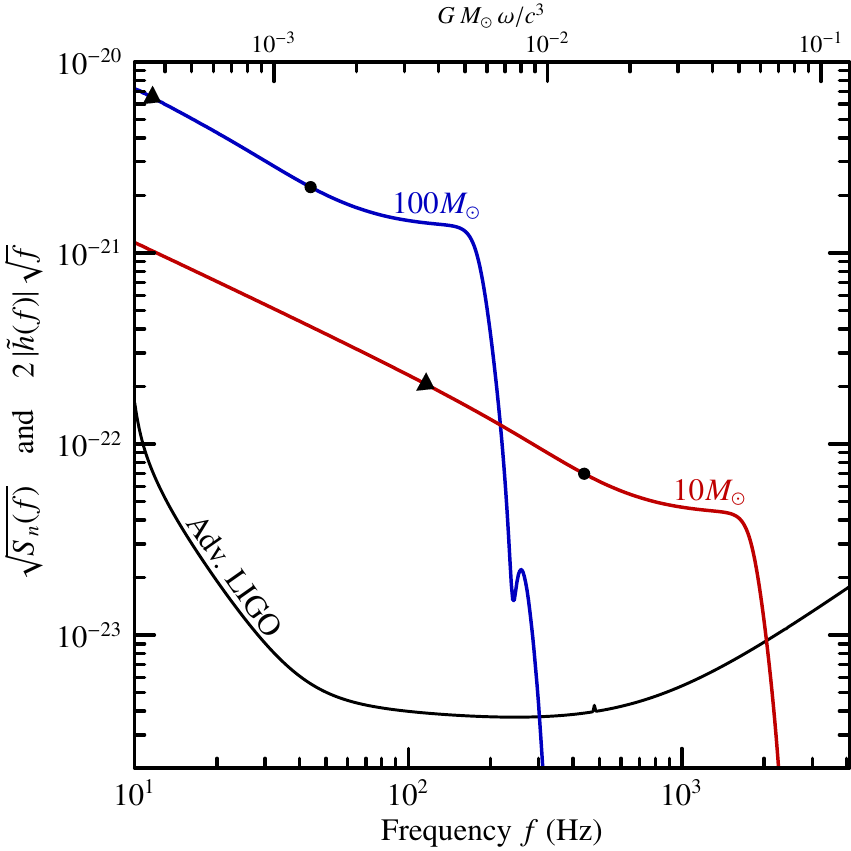}
  \caption{\CapName{Waveforms in the frequency domain} Amplitudes from
    an equal-mass nonspinning binary are shown, scaled to total system
    masses of \SI{10}{\SolarMass} and \SI{100}{\SolarMass} at
    \SI{100}{\mega \parsec}, and compared to the noise spectral
    density of Advanced LIGO (both quantities in units of
    \si[inter-unit-separator=\cdot]{\strain \per \root \hertz}).  The
    factors multiplying the amplitudes are chosen to account for the
    logarithmic scaling of the horizontal axis, the factor of 2 in
    Eq.~\eqref{eq:InnerProduct}, and the fact that only positive
    frequencies are plotted.  The triangles on the waveforms show the
    approximate initial frequency of the longest numerical simulation
    currently available.  The circles show the frequency of the
    innermost stable circular orbit (ISCO)---basically the frequency
    at which PN approximations are expected to be useless.  This plot
    shows that, for the \SI{10}{\SolarMass} system, nearly half the
    contribution to the inner product of Eq.~\eqref{eq:InnerProduct}
    comes from the PN data (to the left of the triangle), and the rest
    from the NR data (to the right).  For the \SI{100}{\SolarMass}
    system, on the other hand, the inner product is given almost
    exclusively by NR data.  \label{fig:SignalsAndNoise}}
\end{figure}

In constructing of a range of plausible model waveforms, we need to
understand the ultimate form of measurement when designing templates
for gravitational-wave detection: the match.  This quantity is based
on the inner product between two waveforms defined as the integral of
the noise-weighted product of the signals in the frequency
domain~\cite{Finn:1992}:
\begin{subequations}
  \label{eq:InnerProduct}
  \begin{align}
    \label{eq:InnerProductWithoutCosine}
    \InnerProduct{h_{a} | h_{b}} &= 2\, \Re \int_{-\infty}^{\infty}
    \frac{\htilde_{a} (f)\, \htilde_{b}^{\ast}(f)}
    {S_{n}(\abs{f})}\, df \\
    \label{eq:InnerProductWithCosine}
    &= 4\, \int_{0}^{\infty} \frac{\abs{\htilde_{a} (f)}\,
      \abs{\htilde_{b}^{\ast}(f)}} {S_{n}(\abs{f})}\, \left[ \cos
      \delta \phi(f) \right]\, df~,
  \end{align}
\end{subequations}
where $S_{n}(f)$ is the power spectral density (PSD) of noise in the
detector, and $\delta \phi(f)$ is the phase difference between the two
waveforms in the frequency domain.  In the following, we will assume
$f>0$.  The second form shown here demonstrates a useful way to
understand the inner product, by separating the integrand into two
factors.  We begin with the first factor, the ratio of amplitudes to
noise.  The amplitude for an example system and the Advanced LIGO
noise curve are plotted in Fig.~\ref{fig:SignalsAndNoise}.  The height
of the amplitude above the noise curve shows when the inner product
can rapidly accumulate; more height means more rapid contribution to
the inner product.  Of course, rapid contributions cannot result in a
large inner product unless those contributions are also coherent.
This requires the second factor of the integrand in
Eq.~\eqref{eq:InnerProductWithCosine} to be constant.  In particular,
a large inner product requires $\delta \phi$ to be very close to zero
across the full range of frequencies for which the signal amplitude is
significantly larger than the noise.

While the amplitude of the waveform from a given system is fixed,
$\delta \phi$ has two inherent degrees of freedom we can adjust to
improve the inner product.  These degrees of freedom are related to
the fact that the merger time and orientation of an astrophysical
binary are unknowns that must simply be measured.  Specifically, we
are free to shift the time and phase of either time-domain waveform by
$\Delta T$ and $\Delta \Phi$.  In the frequency domain, this has no
effect on the amplitude of the waveform (the curves of
Fig.~\ref{fig:SignalsAndNoise} will not be affected), but $\delta
\phi$ changes roughly as
\begin{equation}
  \label{eq:PhaseDifferenceChange}
  \delta \phi(f) \to \delta \phi(f) + 2\, \Delta \Phi + 2\, \pi\, f\,
  \Delta T~,
\end{equation}
We can use this time- and phase-shift freedom to ensure that the phase
difference between two waveforms is smallest at frequencies for which
the detector is most sensitive to that particular system, and thus
maximize the inner product.  The maximum possible (normalized) inner
product~\cite{Finn:1992} is called the match, which will be our basic
measure of error:
\begin{equation}
  \label{eq:Overlap}
  \Overlap{h_{a} | h_{b}} = \max_{\Delta T,\, \Delta \Phi}
  \frac{\InnerProduct{h_{a} | h_{b}}} {\sqrt{ \InnerProduct{h_{a} |
        h_{a}} \InnerProduct{h_{b} | h_{b}} }}~.
\end{equation}
This quantity takes a value between \num{0} (for completely dissimilar
waveforms) and \num{1} (for identical waveforms).  Because many of the
matches encountered below will be very close to \num{1}, it is
preferable to use another quantity called the
mismatch~\cite{SathyaprakashDhurandhar:1991, BalasubramanianEtAl:1996,
  Owen:1996}, which is given by
\begin{equation}
  \label{eq:Mismatch}
  \MM{h_{a}}{h_{b}} \define 1 - \Overlap{h_{a} | h_{b}}~.
\end{equation}
Here, values close to \num{0} indicate the waveforms are similar.  The
maximum possible signal-to-noise ratio (SNR) at which a given signal
$s$ can be detected is given by
\begin{equation}
  \label{eq:MaxSNR}
  \rho_{s} \define \sqrt{ \InnerProduct{s | s} }~.
\end{equation}
The mismatch $\MM{s}{h}$ between a signal $s$ and a template $h$ is
essentially the percentage loss in SNR due to errors in the
template~\cite{Lindblom:2008}.

This understanding of the match teaches us two very important lessons
that will guide our approach to our problem.  First, the optimal phase
shift---and thus the value of the match---will depend on the relative
distribution of power at different frequency bands.  If, for example,
our model waveform simply ends at ISCO, it will fail to model the a
substantial portion of the physical waveform.  In that case, the
maximization of Eq.~\eqref{eq:Overlap} will not need to balance the
dephasing between the two portions of the waveform, for example.  The
requirements for phase coherence will be much looser than they, in
fact, need to be.  Thus, our model waveforms must have roughly the
same distribution of amplitude as the physical waveform, across the
entire sensitive frequency band of the detector.  If our objective is
to evaluate matches before any numerical simulation is done, we will
need a suitable approximation to the merger/ringdown waveform.  For
that purpose, this paper uses the EOB
waveform~\cite{BuonannoDamour:1999, BuonannoDamour:2000,
  BuonannoEtAl:2009a, PanEtAl:2010}, which extends through merger to
ringdown.  Other complete waveforms could also be
used~\cite{AjithEtAl:2007, AjithEtAl:2009, SturaniEtAl:2010}.
Fortunately, we will see that the final results will not depend
strongly on the choice of ersatz NR waveform.  For example, after a
simulation is done, we can go back and check that the results agree if
we use the NR waveform itself.  This is done in
Appendix~\ref{sec:ImportanceOfErsatzNRWaveform} for the equal-mass
nonspinning system.  Even more extreme, the original
proof-of-principle demonstration of this method~\cite{Boyle:2010} used
stationary-phase approximated (SPA) waveforms terminated at the
light-ring frequency.  For small mismatches, the results achieved in
that test compare well to the results achieved in this paper, even
though the SPA waveform mismatches the numerical and EOB waveforms by
more than \SI{8}{\percent} over the relevant mass
range~\cite{BoyleEtAl:2009}.

The second lesson gleaned from these considerations of the match is
that the \emph{phase} of a model waveform does not come into the
match; only the \emph{phase difference} between models matters, as
shown explicitly in Eq.~\eqref{eq:InnerProductWithCosine}.  This fine
distinction has real importance for us because it implies that the
phase error in our ersatz NR waveform relative to the correct physical
waveform is not important.  Certainly the final model waveform should
resemble the physical waveform as closely as possible, but we will
assume that errors in the portion of the final waveform covered by NR
data will be accounted for separately in the error budget---or are
essentially negligible compared to the PN errors.\footnote{Of course,
  this method readily applies to quite general error budgets.  For
  example, if the expected uncertainty in the numerical waveform can
  be estimated, even if that error depends on the initial frequency of
  the simulation, the error budget can be trivially extended to
  include that estimate.  See
  Sec.~\ref{sec:TargetAccuracyAsFunctionOfHybridFreq} for more
  details.}  %
When comparing two plausible waveforms hybridized with the same ersatz
NR data at some frequency $\fhyb$, the phase difference during the NR
portion ($f > \fhyb$) will be zero to a very good approximation, at
least until the waveforms are shifted in time and phase to maximize
the match.  But even then, the phase difference will not depend in any
way on the phase of the ersatz NR waveform; by
Eq.~\eqref{eq:PhaseDifferenceChange}, it will be
\begin{equation}
  \label{eq:PhaseDifferenceErsatzNR}
  \delta \phi(f) = 2\, \Delta \Phi + 2\, \pi\, f\, \Delta T \quad
  \text{for $f > \fhyb$.}
\end{equation}
Thus, the mismatch between plausible hybrids is not directly sensitive
to the particular phasing of the ersatz NR waveform.  Of course, that
phasing will affect the alignment during hybridization, which can
affect the relative power in different portions of the waveform or the
function $\delta \phi(f)$ for frequencies $f < \fhyb$.  However, the
results below show that the ersatz waveform is not dominating the
uncertainty, which suggests that even this phase error is not
important.

These considerations lead us to conclude that the ersatz NR waveform
must be reasonably accurate, especially in terms of modeling the
relative power in different parts of the waveform.  However, we are
also given hope that our final result---the predicted value for
$\Omega_{0}$---does not depend strongly on the accuracy of that ersatz
NR data.  We can test this expectation and will see in
Appendix~\ref{sec:ImportanceOfErsatzNRWaveform} that, at least in the
case of the equal-mass nonspinning system, it is indeed well founded.

\subsection{The range of plausible PN waveforms}
\label{sec:rangeplausiblewaveforms}
Reliable final results depend on an accurate range of plausible
hybrids, correctly depicting the possible error in the analytical
models.  The range must be neither too broad nor too narrow: too
broad, and we will conclude that the error is large, and thus begin
the simulation earlier than necessary; too narrow, and we will be
overconfident in our models, and thus waste time producing an
inaccurate hybrid waveform.  Unfortunately, we have no way of knowing
the error in our analytical models before the fact.  If, however, we
assume that our models are not wrong, but are simply incomplete, we
should be able to trust the uncertainties in the model to estimate the
error.

Analytical relativity has produced multiple methods of calculating
waveforms for black-hole binaries.  Roughly speaking, these different
methods should be equivalent at the level of our knowledge of the true
waveform.  To the extent that they are different, they are uncertain.
In fact, we will use precisely this range of differences as our range
of uncertainty, and thus our estimate for the errors in the analytical
waveforms.  Thus, choosing our range of plausible hybrids comes down
to choosing representatives of the various methods for calculating
analytical waveforms.  The representative methods of calculation we
will use are the TaylorT1--T4~\cite{DamourEtAl:2001,
  BuonannoEtAl:2007, BoyleEtAl:2007} and
EOB~\cite{BuonannoDamour:1999, PanEtAl:2011} models.

An objection might be raised that the EOB waveform is more accurate
and, in particular, ``breaks down'' more elegantly than the basic PN
waveforms; the implication being that the TaylorT$n$ waveforms should
not be included.  It would certainly be possible to use EOB waveforms
alone, employing so-called flexibility
parameters~\cite{DamourEtAl:2003} to delimit a range of plausible
waveforms, for example.  Unfortunately, while EOB waveforms can be
tuned very precisely to resemble the late-time behavior of numerical
waveforms after the fact~\cite{MroueEtAl:2008, PanEtAl:2011}, there is
no evidence that the \emph{inspiral}---which is of more interest
here---will be more accurate after this
tuning~\cite{BrownSathyaprakash:2010}, or that any portion of the EOB
waveforms will be more accurate before such tuning can be done.  In
fact, Blanchet has suggested that EOB appears to converge toward a
theory which is different from general
relativity~\cite{Blanchet:2006}.  Because the results will need to
apply in regions of parameter space where no simulation has yet been
done, and because this paper attempts to reflect methods currently in
use by the community~\cite{NINJA2:2011}, we will take the more
conservative approach of including the TaylorT$n$ approximants.

\section{Evaluating uncertainty}
\label{sec:EvaluatingUncertainty}

With a selection of plausible waveforms in hand, we can now evaluate
the differences between them in terms of the match.  The aim is to
treat the artificial data (the ersatz NR waveform) exactly as data
from the actual simulation will be treated.  First, methods of
hybridization will be reviewed.  This hybridization will be performed
for a variety of hybridization frequencies, \omegah.  Then, we can
simply evaluate the match between the various hybrids at each \omegah,
as a function of the total mass of the system.

\subsection{Hybridization techniques}
\label{sec:Hybridizationtechniques}
Combining inspiral and merger/ringdown waveforms is a delicate
process, beginning with the procedure for aligning the waveforms by
matching up the arbitrary time and phase offsets in the data.  As
described by MacDonald et al.~\cite{MacDonaldEtAl:2011}, this part of
the process has large potential effects on the accuracy of the final
result; in their example a misalignment of just \SI{1}{\Mass} in the
time values of the two waveforms resulted in a mismatch of up to
$\Mismatch = \num{0.01}$.  Clearly, this part of the process must be
handled carefully.  Many techniques have been devised for doing so,
resulting in a variety of choices to be made.

First, alignment of the time and phase offsets may be done in either
the time domain or the frequency domain.  For the particular case of
hybridizing to numerical waveforms, the numerical data is short,
beginning at high frequency and---in particular---having large
amplitude.  Transforming such data into the time domain will either
introduce Gibbs phenomena, which will spoil much of the NR data, or
require windowing, which will waste much of the NR data.  Therefore,
we use time-domain alignment for our purposes, as is used throughout
most of the current literature.\footnote{In the original
  proof-of-principle demonstration of the method described in this
  paper~\cite{Boyle:2010}, frequency-domain alignment was used.  In
  that case, the ersatz NR and PN waveforms were various versions of
  the TaylorF2 waveform---which is calculated in the frequency domain,
  which means Gibbs phenomena are not relevant.  See also
  Ref.~\cite{SantamariaEtAl:2010}.}  %

Second, we must choose a criterion for deciding how well the two
waveforms are aligned after offsetting the time and phase.  Many
possibilities have been suggested for this purpose, including the
magnitude of the difference in the complex $h$
data~\cite{AjithEtAl:2007}; the gravitational-wave phase and
frequency~\cite{BoyleEtAl:2007}; even the \emph{orbital} phase and
frequency~\cite{BuonannoEtAl:2007}.  The gravitational-wave phase and
frequency are often chosen by numerical-relativity groups to produce
hybrids because the post-Newtonian phase is known to higher order than
the amplitude, and is thus more likely to result in an accurate
alignment.  Because of its popularity and simplicity, phase alignment
is used in this paper.

Finally, the alignment procedure depends on the width of the region
over which the criterion chosen above is evaluated.  For example, to
align the phase (and implicitly frequency) of two waveforms, a common
method~\cite{BoyleEtAl:2009} is to minimize the squared difference
between them:
\begin{equation}
  \label{eq:MatchingPhaseOverRegion}
  \Xi(\Delta T, \Delta\phi) = \int_{t_1}^{t_2} \left[
    \phi_{\text{NR}}(t)- \phi_{\text{PN}}(t+\Delta T) - \Delta\phi
  \right]^2 \, dt~.
\end{equation}
The alignment then depends on $t_{1}$ and $t_{2}$.  There is a certain
trade-off here, between using a short region so that less of the
inaccurate PN waveform can be used, and using a long region to smooth
out any irregularities in the numerical data, such as junk radiation
or residual eccentricity.  Additionally, the range $[t_{1}, t_{2}]$
must capture some curvature in the graph of $\phi(t)$ for an accurate
alignment, which means that the range must become larger at lower
frequencies.  Reference~\cite{MacDonaldEtAl:2011} suggests a simple
but robust method of choosing this range, where $t_{1}$ and $t_{2}$
extend to frequencies \SI{5}{\percent} above and below some central
frequency.  This ensures that the range is neither too large at high
frequencies, nor too small at low frequencies.

In our case, the ``numerical'' data is the EOB waveform, which has
essentially no eccentricity or noise.  Thus, for simplicity of
presentation and implementation, we take the limit of this procedure
as $t_{2}$ approaches $t_{1}$.\footnote{It has been checked that the
  results of this paper are essentially identical when using
  Eq.~\eqref{eq:MatchingPhaseOverRegion} with values of $t_{1}$ and
  $t_{2}$ as prescribed in Ref.~\cite{MacDonaldEtAl:2011}.  This is a
  better choice for data from simulations, and is more typical of
  hybridization as practiced by numerical-relativity groups, but
  would introduce an unnecessary layer of complexity to the discussion
  here.}  %
To do so stably, we adjust the time offset $\Delta T$ so that the
frequencies are the same at some time $\thyb$, then adjust the phase
offset $\Delta \phi$ so that the phases are the same at $\thyb$:
\begin{subequations}
  \label{eq:PhaseAndFrequencyMatching}
  \begin{gather}
    \omega_{\text{NR}}(\thyb) = \omega_{\text{PN}}(\thyb+\Delta T)~,
    \\
    \phi_{\text{NR}}(\thyb) = \phi_{\text{PN}}(\thyb+\Delta T) +
    \Delta \phi~.
  \end{gather}
\end{subequations}
Here, the optimal offsets will depend on the time at which the
alignment condition is imposed.  Below, this dependence is described
using the frequency itself: $\omegah \define \omega_{\text{NR}}
(\thyb)$.

The alignment just described is typically applied to the $(\ell,m) =
(2,2)$ mode in a spin-weighted spherical harmonic decomposition of the
gravitational waves.  Other modes must also be aligned.  However, we
have fixed the only degrees of freedom, which means that the other
modes are already determined.  In general, the amplitude and phase of
any mode of the PN waveform is transformed according to
\begin{subequations}
  \label{eq:HigherModePhaseAndFrequencyMatching}
  \begin{gather}
    A_{\text{PN}}^{\ell,m}(t) \to A_{\text{PN}}^{\ell,m} (t-\Delta
    T)~,
    \\
    \phi_{\text{PN}}^{\ell,m}(t) \to
    \phi_{\text{PN}}^{\ell,m}(t-\Delta T) - m\, \Delta \phi / 2 + 2\,
    \pi\, n~,
  \end{gather}
\end{subequations}
for some integer $n$ that ensures continuity of the phase.

Now, having aligned the two waveforms, we need to produce a single
waveform.  Because only $\phi^{2,2}$ is guaranteed to be continuous,
discontinuities are possible in other quantities, so a hybrid is
usually formed using a transition function to blend the two waveforms.
Here, because the discontinuities are mild, we use a basic linear
transition~\cite{BoyleEtAl:2009} of width \SI{10}{\Mass} centered on
the alignment point, for amplitudes and phases of all modes.

A special case arises when the PN waveform is the EOB approximant.
Rather than actually splitting the EOB waveform into two parts and
recombining them, the complete EOB waveform is used as the EOB
``hybrid''.

\subsection{Mismatches}
\label{sec:Mismatches}
Now, having formed a series of hybrids using various PN approximants,
and a range of hybridization frequencies $\omegah$, we can evaluate
the difference between them using the same criterion as will be used
with the numerical data.  In particular, we take the mismatch
[Eq.~\eqref{eq:Mismatch}] using the Advanced LIGO zero-detuning,
high-power noise curve~\cite{Shoemaker:2010}.  The waveforms are
projected onto the positive $z$ axis using all available modes of the
spin-weighted spherical-harmonic decomposition.  Post-Newtonian
calculations have been carried out through $\ell = 8$:
\begin{equation}
  \label{eq:WaveformAsSumOfModes}
  h(t) = \Re\, \left[ \sum_{\ell=2}^{8}\, \sum_{m=-\ell}^{\ell}\,
    A^{\ell, m}(t)\, \e^{\i\, \phi^{\ell, m}(t)}\, \mTwoYlm{\ell,
      m}(0,0) \right]~.
\end{equation}
Note, however, that the quasinormal-mode portion of the EOB waveform
has only been extended to include $(\ell, m) \in \{(2,\pm 2), (2,\pm
1), (3, \pm 3), (3, \pm 2), (4, \pm 4)\}$~\cite{PanEtAl:2011}.  Other
modes are set to zero during ringdown.

\begin{figure}
  \includegraphics[width=\linewidth]{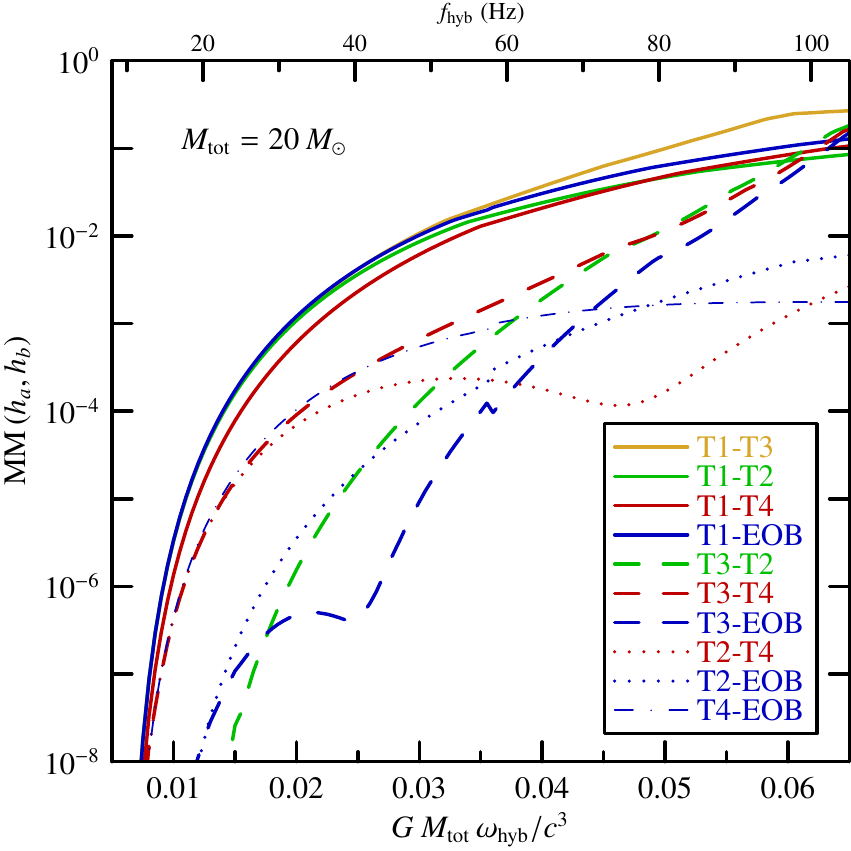}
  \caption{\CapName{Mismatches between hybrids as a function of
      $\omegah$} This plot shows the mismatch between pairs of hybrids
    using different approximants, for the equal-mass nonspinning
    system with $\Mtot = \SI{20}{\SolarMass}$.  At any particular
    value of $\omegah$, the maximum mismatch between each pair of
    hybrids is the uncertainty in the final waveform hybridized at
    that frequency.  If our target accuracy were, for example,
    $\Mismatch \leq \num{e-3}$ for this system mass, this plot shows
    us that the NR waveform would need to contain the GW frequency of
    $G\, \Mtot\, \omegah / c^{3} = \num{0.02}$, which naturally
    implies the initial orbital frequency of the simulation,
    $\Omega_{0}$.  Achieving a smaller uncertainty requires lower
    $\omegah$, corresponding to times at which the PN approximants
    agree more closely.  \label{fig:MismatchVersusOmegah}}
\end{figure}

As a first example, Fig.~\ref{fig:MismatchVersusOmegah} shows the
mismatches between each pair of hybrids for the equal-mass nonspinning
system scaled to a total mass of \SI{20}{\SolarMass}.  At any
particular value of $\omegah$ there is a range of mismatches,
indicating that some pairs of hybrids happen to agree with each other
very closely, while some are quite different.  There is no reason to
suspect that a pair of hybrids in close agreement with each other also
agree with the exact waveform.  Rather, if these are all plausible
waveforms, the uncertainty in our model is given by the maximum
mismatch between any pair.  For this particular system, that pair
happens to be the TaylorT1 and T3 waveforms for most values of
$\omegah$, though in general TaylorT1 is most dissimilar from the
other waveforms.

We can follow the maximum mismatch as a function of frequency, and
notice a general trend: increasing hybridization frequency results in
increasing uncertainty.  This is to be expected for two reasons.
First, the PN approximation should be very accurate at low
frequencies, but break down at higher frequencies.  For example, the
time and phase at which $\omegah$ occurs in the PN waveform will
become more uncertain as $\omegah$ increases.  This results in
uncertainty in the alignment between the two parts of the hybrid.

Second, as the hybridization frequency increases, the detector will
simply be more sensitive to the differences between hybrids.  The
upper horizontal axis of Fig.~\ref{fig:MismatchVersusOmegah} shows the
physical hybridization frequency $\fhyb$.  Comparison with the noise
curve plotted in Fig.~\ref{fig:SignalsAndNoise} demonstrates that the
mismatch grows very quickly just as the hybridization point passes the
``seismic wall'' of the detector ($f_{\text{seismic}} =
\SI{10}{\hertz}$) where the sensitivity is improving rapidly with
increasing frequency.  The mismatch then begins to level out as the
detector sensitivity levels out.

In the next section, we discuss how to use plots like the one in
Fig.~\ref{fig:MismatchVersusOmegah}, along with a target accuracy, to
find the optimal initial orbital frequency of a numerical simulation.

Before moving on, however, let us pause to note an interesting feature
of the last plot.  The various comparisons separate into distinct
groups.  The largest mismatches involve TaylorT1 hybrids (solid
lines), and \emph{all} hybrids using T1 have large mismatches with
other hybrids.  Setting aside the T1 waveforms, we see a similar trend
develop at high frequencies for T3 hybrids compared to hybrids other
than T1 (dashed lines).  We could even push this categorization to the
T2 hybrids compared to waveforms other than T1 and T3 (dotted lines),
though this leaves little room for comparison.  We might say that T1
is dominating the uncertainty, in the sense that it is furthest from
the consensus of other hybrids.  At higher frequencies, T3 departs
from that consensus, followed by T2, leaving only the T4 and EOB
agreeing with each other.  What is striking about this pattern of
\emph{disagreements} is that it is identical to the pattern in
\emph{errors} relative to the numerical
waveform~\cite{BoyleEtAl:2007}, where T1 is least accurate, followed
by T3, with T2 slightly less accurate than T4; the T4 and EOB
waveforms agree with the numerical result nearly within numerical
errors.

An optimist might suggest that close systematic agreement between two
models like T4 and EOB is unlikely---given the size of the function
space through which they are free to roam---unless they also agree
with the exact waveform.  This would imply that we should take the
small mismatch between T4 and EOB as the uncertainty in those
waveforms.  Or, slightly less optimistically, we might discount T1 as
being too far from the other waveforms, and thus a mere anomaly.
Unfortunately, similar patterns do not develop for the other systems
investigated below.  For now, we leave this as a mere observation, and
take the largest mismatch as an indicator of the uncertainty in any
given model.

\section{Using mismatch to find \texorpdfstring{$\Omega_0$}
  {Omega0}}
\label{sec:UsingTheOverlapsToFindOmegaNought}
Given a target accuracy, the uncertainty implied by
Fig.~\ref{fig:MismatchVersusOmegah} suggests a natural starting point
for the numerical simulation of that system, simply because the
simulation must include---at a minimum---the corresponding $\omegah$.
In this section, the uncertainty estimate of the previous section is
used to produce an optimal initial orbital frequency for the
simulation.  This is then generalized to apply across
a range of masses and to incorporate more complicated target
mismatches.  In the process, the uncertainties for a small selection
of astrophysical systems are shown.

\subsection{Optimizing \texorpdfstring{$\Omega_0$}{Omega0} for a
  particular mass}
\label{sec:OptimizingOmega_0ForAParticularMass}

Figure~\ref{fig:MismatchVersusOmegah} establishes a relationship
between the uncertainty in plausible hybrids and the frequency at
which they are hybridized, $\omegah$.  If, for example, we wish to
model an equal-mass nonspinning system of total mass $\Mtot/\MSun =
\num{20}$ with a target accuracy of $\TargetMismatch \leq \num{e-3}$,
this plot demonstrates that the final hybrid waveform must be formed
with $G\, \Mtot\, \omegah / c^{3} \lesssim \num{0.02}$.  Naturally,
the numerical simulation must include that frequency.  There are two
simple methods for turning the GW frequency into an initial orbital
frequency for the simulation, $\Omega_{0}$.

First, we might use the basic approximation $\Omega_{0} \approx
\omegah / 2$.  In this case, the result above would suggest an initial
frequency of $G\, \Mtot\, \Omega_{0} / c^{3} \lesssim
\num{0.01}$.\footnote{This is roughly half the initial frequency of
  current long simulations.  To lowest order, the length of a
  simulation goes as $T \propto \Omega_{0}^{-8/3}$, which means that a
  simulation held to this standard needs to be roughly $2^{8/3}
  \approx \num{6.3}$ times longer than current long simulations.}  %
The actual simulation should probably begin somewhat earlier than
this, to allow junk radiation to leave the system, and to ensure that
the alignment region of Eq.~\eqref{eq:MatchingPhaseOverRegion} does
not extend past $\omegah$.  These considerations depend on the
particular formulation of Einstein's equations and numerical methods
used in the simulation, and are thus beyond our scope.  Ultimately,
the numerical relativist will use his or her judgment to produce some
time $\Delta t$ before $\omegah / 2$ at which the simulation should
begin.  In this regard, an additional PN approximation may be useful:
\begin{equation}
  \label{eq:OmegaNoughtOfomegah}
  \Omega_{0} \approx \frac{\omegah} {2} -\Delta t\, \left(
    \frac{\omegah} {2} \right)^{11/3}\, \frac{96\, \nu} {5}\, \left(
    \frac{G\, \Mtot} {c^{3}} \right)^{5/3}~,
\end{equation}
where $\nu = m_{1}\, m_{2} / (m_{1}+m_{2})^{2}$ is the symmetric mass
ratio of the individual black holes.  The extra term is derived from
the lowest-order PN approximation for the evolution of the orbital
frequency~\cite{Blanchet:2006}.  Because of the approximations, this
method may fail in certain extreme cases.

Alternatively, and more robustly, we might simply refer to any of the
PN models contributing to our estimate, to find the orbital frequency
corresponding to $\omegah$.  Moreover, if some $\Delta t$ is
prescribed for the simulation, the PN model can be used to find the
orbital frequency $\Omega_{0}$ occurring at a time $\Delta t$ before
the GW frequency $\omegah$.

The particular example just discussed applies only when the target
accuracy is required for $\Mtot = \SI{20}{\SolarMass}$.  This would be
relevant to the situation where, for example, a source has been
detected, and its parameters are known to reasonable accuracy, but
further simulations are being done for accurate parameter estimation.
More generally, however, we should expect to encounter broader
accuracy requirements, which might apply across a range of
masses~\cite{NRAR:2011}.  The rest of this section will extend
this example to account for various masses; to demonstrate the
uncertainty for a selection of interesting systems; then to allow the
target accuracy to vary as a function of mass; and finally to allow
the target accuracy to vary as a function of both mass and
hybridization frequency.

\subsection{Optimizing \texorpdfstring{$\Omega_{0}$}{Omega0} for a
  range of masses}
\label{sec:ConstantTargetAccuracy}
\begin{figure*}
  \includegraphics[width=\columnwidth]%
  {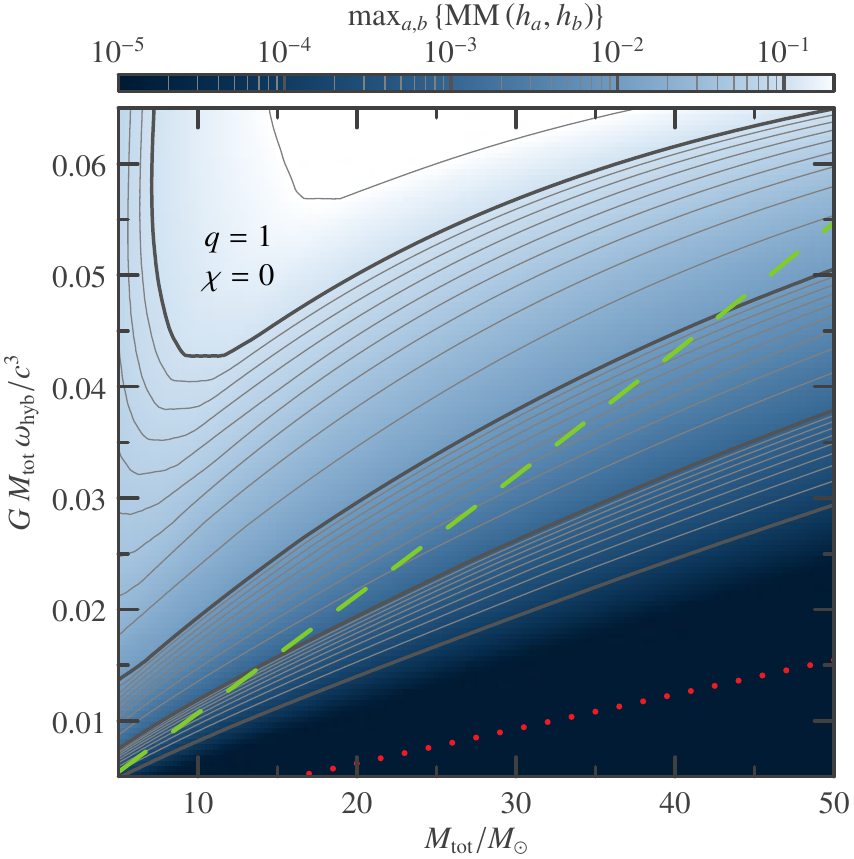} \hfill
  \includegraphics[width=\columnwidth]%
  {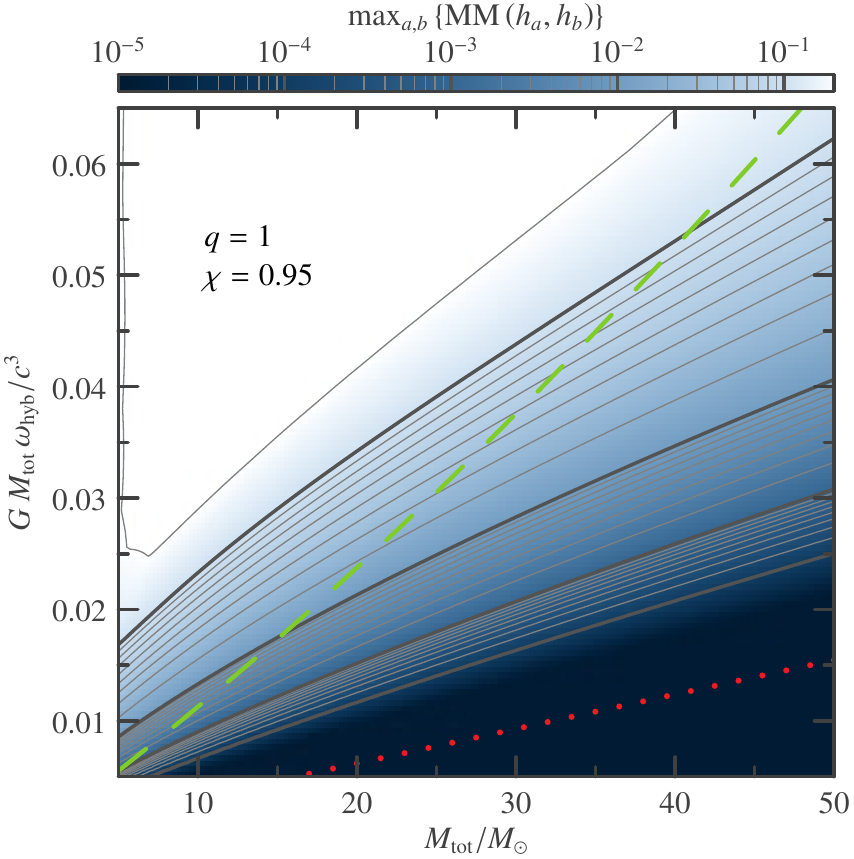} \\[20pt]
  \includegraphics[width=\columnwidth]%
  {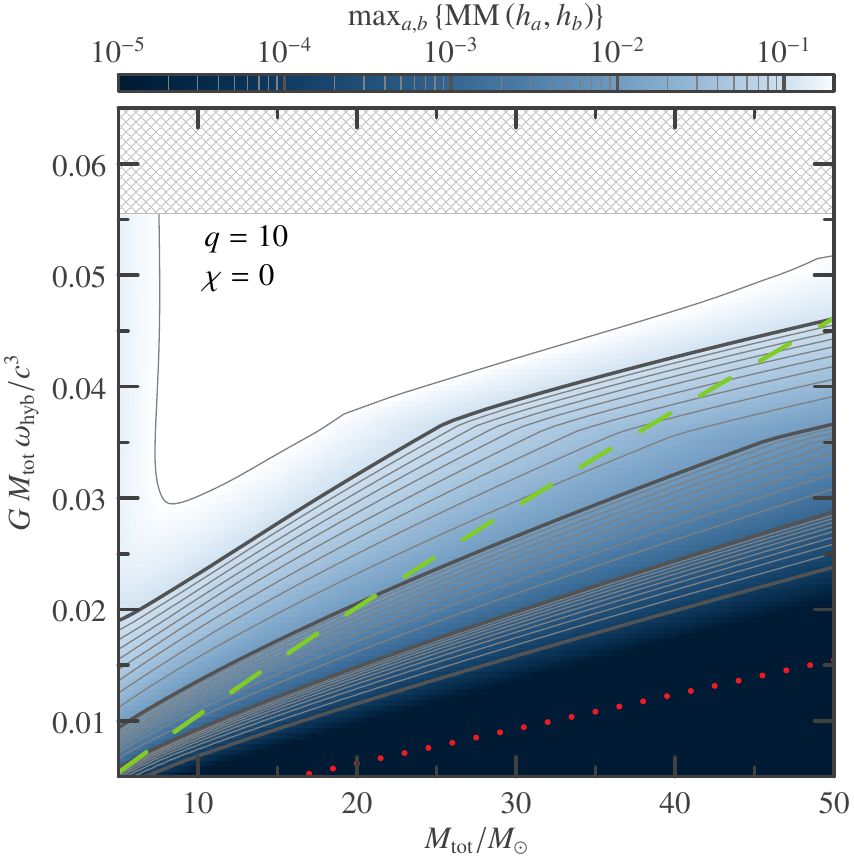} \hfill
  \includegraphics[width=\columnwidth]%
  {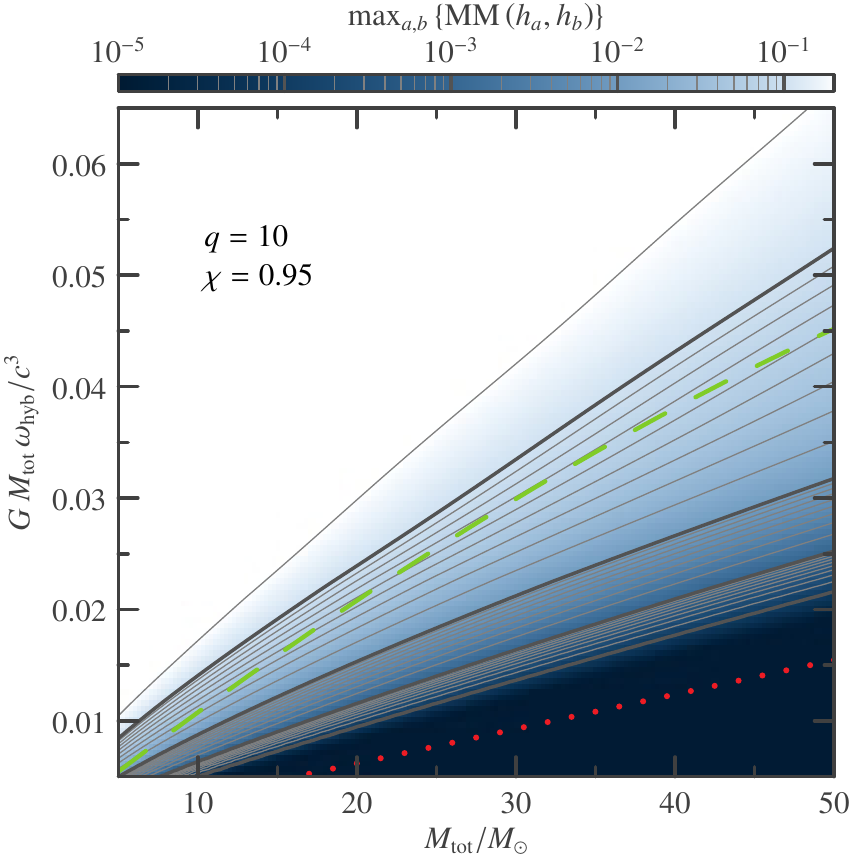}
  \caption{\CapName{Maximum mismatch between plausible hybrids for a
      selection of systems} These plots show the maximum mismatch
    between any pair of hybrids formed by hybridizing TaylorT1--T4 or
    EOB with an ersatz NR waveform (EOB) at $\omegah$, as described in
    Sec.~\ref{sec:Hybridizationtechniques}, when scaled to a total
    system mass of $\Mtot$.  This quantity describes the uncertainty
    in the models used to form the hybrid.  The plots show data from
    four cases with mass ratios denoted by $q$, which is the ratio of
    the larger to smaller mass, and components of the dimensionless
    spins $\chi$ aligned with the orbital angular-momentum vector.
    Note that $\Mtot \lesssim \SI{33}{\SolarMass}$ may not be
    interesting astrophysically for black-hole binaries when $q =
    \num{10}$.  The region $G\, \Mtot\, \omegah / c^{3} \gtrsim
    \num{0.055}$ is inaccessible in the $q = \num{10}$, $\chi =
    \num{0}$ case, because the TaylorT3 approximant ends at that
    frequency; this could be interpreted as complete uncertainty.
    Generally, the uncertainty is larger in systems with more extreme
    parameters.  The dotted red line in each plot shows
    $f_{\text{seismic}} = \SI{10}{\hertz}$, the lower bound of
    sensitivity in Advanced LIGO.  The dashed green line in each plot
    shows $f_{\SI{20}{\percent}}$---the twentieth percentile of the
    power of the match, defined by Eq.~\eqref{eq:Definefpercentile}.
    Comparing any plot to an accuracy requirement, which may depend on
    both $\Mtot$ and $\omegah$, we can extract the maximum sufficient
    hybridization frequency, which suggests the optimal initial
    orbital frequency.  The method for doing this is described in
    Sec.~\ref{sec:UsingTheOverlapsToFindOmegaNought}.  These plots are
    discussed in more detail in
    Sec.~\ref{sec:TheCornersOfParameterSpace}.
    \label{fig:MismatchContours}}
  \vfill
\end{figure*}

The mismatch curves plotted in Fig.~\ref{fig:MismatchVersusOmegah}
depend strongly on our choice of the total system mass.  To generalize
this to be a function of both $\omegah$ and $\Mtot$, we create the
contour plot in the upper left of Fig.~\ref{fig:MismatchContours}.
Slicing through that plot at $\Mtot / \MSun = \num{20}$ gives the
uppermost curve in Fig.~\ref{fig:MismatchVersusOmegah}.  For
comparison, this quantity is also plotted for a selection of systems
with different mass ratios or spins, as discussed in greater detail in
Sec.~\ref{sec:TheCornersOfParameterSpace}.

Again, given an accuracy requirement, we can use this plot to derive
the optimal initial orbital frequency.  If the requirement is a target
accuracy of $\TargetMismatch \leq \num{e-3}$, we can follow the
$\num{e-3}$ contour in the plot, and see that it is always above $G\,
\Mtot\, \omegah / c^{3} \approx \num{0.0075}$ for the range of masses
shown.  As before, the initial orbital frequency $\Omega_{0}$ is then
deduced from this value by using Eq.~\eqref{eq:OmegaNoughtOfomegah} or
by consulting the PN model, as described above.  This stringent
accuracy requirement calls for numerical simulations roughly \num{87}
times longer than the longest current simulation of this system.

\subsection{Comparing the uncertainty in various systems}
\label{sec:TheCornersOfParameterSpace}

While the equal-mass nonspinning case nicely illustrates the method of
finding optimal initial frequencies, systems at the boundaries of
current numerical capabilities also hold a great deal of interest.
Fig.~\ref{fig:MismatchContours} illustrates the uncertainty for four
systems:
\begin{enumerate}
 \item Equal-mass, nonspinning;
 \item Equal-mass, aligned spins $\chi_{1} = \chi_{2} = \num{0.95}$;
 \item Mass ratio \num{10}:\num{1}, nonspinning;
 \item Mass ratio \num{10}:\num{1}, aligned spins $\chi_{1} = \chi_{2}
  = \num{0.95}$.
\end{enumerate}
The quantities $\chi_{1}$ and $\chi_{2}$ are the components of the
dimensionless spins along the orbital angular-momentum vector.  In the
lower-left plot, case (3), frequencies above $G\, \Mtot\, \omegah /
c^{3} \approx \num{0.055}$ are not possible because the TaylorT3
approximant ends at that frequency for this system.  We can treat
those higher frequencies as having $\Mismatch = \num{1}$.  Note that
the smallest known black holes have masses of at least $M \approx
\SI{3}{\SolarMass}$~\cite{GelinoHarrison:2003}.  This suggests that
the smallest total mass for systems with $q=\num{10}$ would be $\Mtot
\approx \SI{33}{\SolarMass}$.  Thus, the low-mass regions of those two
plots may not be interesting astrophysically.

In each case, we see the basic trends noted in
Sec.~\ref{sec:Mismatches}.  Two factors drive the mismatch: how far
$\omegah$ has entered the sensitive band of the detector, and how
poorly the post-Newtonian approximation performs up to that frequency.
Each plot of Fig.~\ref{fig:MismatchContours} includes a dotted red
line denoting $f_{\text{seismic}}$, the lower bound of sensitivity in
Advanced LIGO.  Below this line, the mismatch must be zero, because
the data in the detector's sensitive band is identical for any two
waveforms---it is just the ersatz NR data.  As we move above this
line, a larger fraction of the data in the corresponding hybrids comes
from different approximants.  Thus, the mismatch increases.  We expect
it to increase more quickly, as a function of $\omegah$ for systems
that are not well described by PN approximations.  Indeed, comparing
the plots, we see that any given contour line moves closer to
$f_{\text{seismic}}$ as either the mass ratio or spin parameter
increases.  This is simply an explicit confirmation that the physical
system is not well modeled in more extreme cases.

Though less important for our purposes, the contour lines for larger
mismatches also obey a similar bound.  In each plot, the dashed green
line shows $f_{\SI{20}{\percent}}$, the twentieth percentile of power
in the match.  That is, assuming $\delta \phi = \num{0}$ in
Eq.~\eqref{eq:InnerProductWithCosine}, $f_{\SI{20}{\percent}}$ is the
frequency for which the integral has accumulated \SI{20}{\percent} of
its final value:
\begin{multline}
  \label{eq:Definefpercentile}
  4\, \int_{0}^{f_{\SI{20}{\percent}}}
  \frac{\abs{\htilde_{a} (f)}\, \abs{\htilde_{b}^{\ast}(f)}}
  {S_{n}(\abs{f})}\, df \\= 0.20\times 4\, \int_{0}^{\infty}
  \frac{\abs{\htilde_{a} (f)}\, \abs{\htilde_{b}^{\ast}(f)}}
  {S_{n}(\abs{f})}\, df~.
\end{multline}
For simplicity, the lines shown in the plots use EOB data for both
waveforms: $h_{a} = h_{b} = h_{\text{EOB}}$.  The line is roughly a
lower bound for mismatches of $\Mismatch = \num{0.20}$---the white
region of each plot.  The higher the white region is above the
$f_{\SI{20}{\percent}}$ line, the more accurate the PN approximations
are.  At very high frequencies, the white region must approach this
line simply because the PN approximations break down more quickly
there.

\subsection{Mass-dependent target accuracy}
\label{sec:TargetAccuracyAsFunctionOfMass}

In each of the examples above, the accuracy requirement calls for a
specific mismatch, regardless of the total mass of the system.  This
is unrealistic for three reasons.  First, and most simply, the result
depends sensitively on a choice of mass---in the example of
Sec.~\ref{sec:ConstantTargetAccuracy}, that choice is the lower bound
of the mass range used in Fig.~\ref{fig:MismatchContours}.  If we were
to increase that lower bound to $\Mtot \geq \SI{50}{\SolarMass}$,
rather than calling for simulations \num{87} times longer, we would
accept the longest current simulations.  This sensitivity to the mass
range shows that it must be considered more carefully than an
arbitrary choice of plotting range.

Second, the SNR of an astrophysical signal will depend on the total
mass.  The precise dependence is complicated, as it involves the shape
of the noise curve in Fig.~\ref{fig:SignalsAndNoise}.  However, for
masses at which the merger frequency is much higher than the
detector's low-frequency sensitivity (a good approximation for the
mass range discussed here), the stationary-phase approximation shows
that the SNR should scale roughly as
$\Mtot^{5/6}$~\cite{CutlerFlanagan:1994}.  If we expect the typical
low-mass system in real data to have a lower SNR than the typical
high-mass system, there is no reason to model the two with the same
precision.  More precisely, for optimal parameter estimation, the
error of a model waveform should scale inversely as the square of the
SNR of the expected signal~\cite{FlanaganHughes:1998, Lindblom:2008}.

Finally, the merger rates of real binaries will likely depend on the
total mass simply because formation mechanisms for such binaries
should depend on the total mass~\cite{BulikBelczynski:2003,
  OshaughnessyEtAl:2010, FarrEtAl:2010, OzelEtAl:2010}.  Though
current understanding of such mass dependence is not
great~\cite{LIGO:2010}, it is an area of active research, and will no
doubt be improved in the future.  In that case, we may wish to fold
the expected event rate into our target accuracy, so that time is not
wasted calculating precise waveforms for systems we are unlikely to
observe.

For these reasons, useful and efficient accuracy requirements should
depend explicitly on $\Mtot$.  Incorporating mass dependence involves
reinterpreting the target mismatch $\TargetMismatch$ as a function of
mass, rather than as a constant.  For example, we might hope to model
a particular system well enough to ensure detection of any binary
expected in the data, and to allow accurate parameter estimation for
\SI{90}{\percent} of binaries.  A very crude function that implements
this idea is
\begin{equation}
  \label{eq:TargetMismatchAsFunctionOfMass}
  \TargetMismatch(\Mtot) =
  \begin{cases}
    \num{1} & \frac{\Mtot}{\MSun} \leq \num{7}; \\
    \num{e-2} & \num{7} < \frac{\Mtot}{\MSun} \leq \num{12}; \\
    \num{e-4} \left(\frac{\Mtot} {\SI{20}{\SolarMass}} \right)^{-5/3} &
    \num{12} < \frac{\Mtot}{\MSun}.
  \end{cases}
\end{equation}
The first two cases are inspired by population-synthesis
results~\cite{BulikBelczynski:2003} suggesting that basically all
equal-mass black-hole binaries should have\footnote{Note that several
  of the papers cited here discuss individual black-hole mass or the
  chirp mass of a binary, rather than the total mass.}  %
$\Mtot \gtrsim \SI{7}{\SolarMass}$, and that roughly \SI{90}{\percent}
should have $\Mtot \gtrsim \SI{12}{\SolarMass}$.  This function
ignores binaries with $\Mtot \leq \SI{7}{\SolarMass}$, and models
binaries with $\Mtot \leq \SI{12}{\SolarMass}$ just well enough to
ensure detection.\footnote{A mismatch of $\num{e-2}$ or less
  \emph{ensures} detection \SI{97}{\percent} of the time for a signal
  with SNR at least \num{7}~\cite{FlanaganHughes:1998}.  However, a
  real search of detector data will use a template bank which does not
  need to be as accurate as this, in the sense that an inaccurate
  template with different physical parameters may happen to match the
  exact waveform.  That is, the error of an ``effectual'' template
  bank may be far greater than $\num{e-2}$.  The issue of detection is
  complicated, and is discussed at length in
  Ref.~\cite{BuonannoEtAl:2009}.}  %
For higher masses, the function scales inversely as the SNR, and is
normalized to optimally estimate the parameters of a
\SI{20}{\SolarMass} system having SNR \num{70}.  Clearly, more
sophisticated treatments could incorporate the objectives of detection
and parameter estimation more smoothly, but this will serve to
illustrate the idea.

\begin{figure}
  \includegraphics[width=\linewidth]%
  {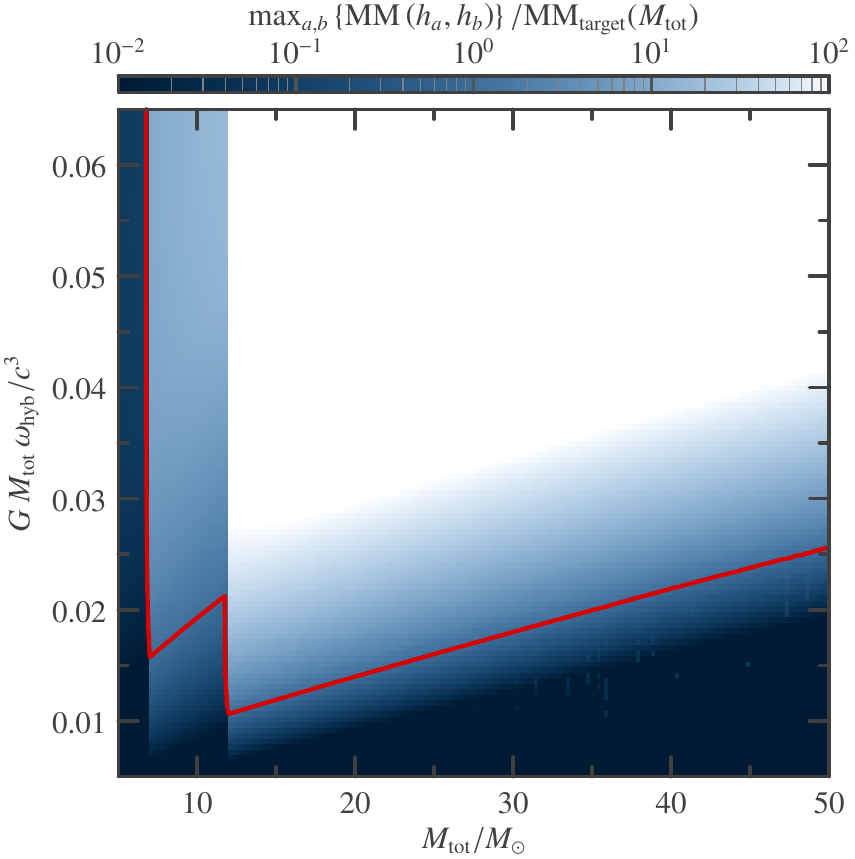}
  \caption{\CapName{Ratio of maximum mismatch to target mismatch} This
    plot shows the ratio between the maximum mismatch between various
    hybrids and a target mismatch given by
    Eq.~\eqref{eq:TargetMismatchAsFunctionOfMass}.  For values of this
    ratio greater than \num{1}, the hybridization frequency is too
    high to achieve the target accuracy.  The optimal sufficient value
    of $\omegah$ is given by the lowest frequency at which the ratio
    is \num{1}---roughly $G\, \Mtot\, \omegah / c^{3} = \num{0.011}$
    here.  \label{fig:TargetMismatchRatio_q1_chis0}}
\end{figure}

Regardless of the particular form of the target mismatch, we use it to
deduce the sufficient value of $\omegah$ (and hence $\Omega_{0}$) by
plotting the ratio
\begin{equation}
  \label{eq:MassDependentRatio}
  \frac{ \max_{a,b} \left\{ \MM{h_a}{h_b} \right\} } {
    \TargetMismatch(\Mtot) }~.
\end{equation}
Where this ratio exceeds \num{1}, the hybridization frequency is
insufficient.  With the target mismatch of
Eq.~\eqref{eq:TargetMismatchAsFunctionOfMass}, this ratio is plotted
in Fig.~\ref{fig:TargetMismatchRatio_q1_chis0}.
The red curve denotes values for which the ratio equals 1.  The
optimal $\omegah$ is given by the lowest point this curve reaches,
$G\, \Mtot\, \omegah / c^{3} \approx \num{0.011}$.  Though this
frequency is only \SI{47}{\percent} higher than the frequency deduced
in the previous section without incorporating mass dependence, it
corresponds to a simulation that is nearly \num{3} times shorter---a
significant improvement from the perspective of the numerical
relativist, and hopefully a more accurate representation of the
accuracy truly required for gravitational-wave detector data analysis.
This holds great significance for systems with large mass ratios,
where astrophysical considerations suggest that the lower bound of
$\Mtot$ will be large, as mentioned in
Sec.~\ref{sec:TheCornersOfParameterSpace}.

\subsection{Mass- and frequency-dependent target accuracy}
\label{sec:TargetAccuracyAsFunctionOfHybridFreq}

Up to this point, we have assumed that the mismatch between our
hybrids completely describes the uncertainty in the final result,
after a numerical simulation has been done.  However, for the
stringent accuracy requirements quoted above, the numerical simulation
will be very long, making it difficult to achieve high accuracy in the
NR portion of the waveform.  We may wish to leave a portion of the
error budget for the NR data, and the rest for the PN data and hybrid.
Given some understanding of how the NR error depends on the length of
the simulation, we can incorporate that error into our determination
of the optimal value of $\Omega_{0}$.  The error of a simulation
depends, no doubt, on its length, but also on the particular
implementation used for that simulation and even the computational
resources available, and is thus beyond the scope of this paper.
However, the basic idea is a simple extension of the technique
discussed in the previous section: generalize the target mismatch to
be a function of both $\Mtot$ and $\omegah$, and plot the ratio
\begin{equation}
  \label{eq:MassAndFreqDependentRatio}
  \frac{ \max_{a,b} \left\{ \MM{h_a}{h_b} \right\} } {
    \TargetMismatch(\Mtot, \omegah) }~.
\end{equation}
Again, when this is greater than \num{1}, $\omegah$ is too large.  For
example, the target mismatch may be constructed (crudely) by setting
an astrophysically motivated target $\Mismatch_{\text{t,astro}}$ for
the final waveform, then subtracting the estimated uncertainty due to
the NR data.  Then the permissible mismatch in the PN data could be
defined as
\begin{multline}
  \label{eq:TargetWithAstroAndNRContributions}
  \TargetMismatch(\Mtot, \omegah) \\ \define
  \Mismatch_{\text{t,astro}}(\Mtot) - \Mismatch_{\text{NR}}(\Mtot,
  \omegah)~.
\end{multline}
This target is crude because the mismatch is not additive, but it is a
conservative estimate.

This extension of the method to include dependence on $\omegah$ raises
an unfortunate---though realistic---possibility: it could be that the
ratio of Eq.~\eqref{eq:MassAndFreqDependentRatio} will never be less
than \num{1} for masses of interest.  For example, if
$\Mismatch_{\text{NR}}(\Mtot, \omegah)$ increases too quickly as
$\omegah$ decreases, the quantity in
Eq.~\eqref{eq:TargetWithAstroAndNRContributions} will be too small,
and thus the ratio in expression~\eqref{eq:MassAndFreqDependentRatio}
will be too large.  This would indicate that the modeling methods,
both analytical and numerical, are simply too crude to compute the
waveform with the desired accuracy.

\section{Discussion}
\label{sec:Conclusions}

Figure~\ref{fig:MismatchContours} presents the main results of this
paper, showing the largest mismatch between any pair of plausible
hybrids as a function of the frequency of hybridization and the total
mass of the system.  The hybrids are formed using the EOB waveform to
substitute for the NR waveform.  As is argued in
Sec.~\ref{sec:Motivation} and shown explicitly for the equal-mass
nonspinning system in Appendix~\ref{sec:ImportanceOfErsatzNRWaveform},
the particular choice of substitute does not affect the final results
in any significant way.  The hybrids' inspiral data are supplied by
TaylorT1--T4 and EOB waveforms, which are attached at $\omegah$.
These hybrids are then scaled to various total masses, and mismatches
between each pair are calculated.  The maximum such mismatch is the
estimated uncertainty in the models.  The plots in
Fig.~\ref{fig:MismatchContours} assume that the error in the numerical
portion of the hybrid is negligible, though they can be expanded to
account for estimated numerical errors, as in
Sec.~\ref{sec:TargetAccuracyAsFunctionOfHybridFreq}.  This uncertainty
is a reasonable proxy for the error---the difference between the model
and the exact waveform.  Given a target uncertainty for the complete
model, we can deduce the minimum initial orbital frequency necessary
to achieve that target with a simulation, by noting that the relevant
value of $\omegah$ must be present in the simulation data.

The results show several interesting features.  First, the uncertainty
generally increases as the modeled system becomes more extreme; for a
given value of $\omegah$ and $\Mtot$, increasing either the mass ratio
or the spin parameter increases the uncertainty.  This is not
surprising, since the post-Newtonian order of known spin terms is
lower than the order for non-spin terms~\cite{Arun:2009}.  Similarly,
PN methods are expected to break down for larger mass
ratios~\cite{Blanchet:2006}, for which more specific methods are
necessary~\cite{SasakiTagoshi:2003}.

More quantitatively, we can relate these results to basic accuracy
standards for gravitational-wave detectors.  To calibrate waveforms
for detection, accuracies of $\Mismatch \lesssim \num{0.01}$ are
generally called for~\cite{FlanaganHughes:1998, Lindblom:2008}.
Meanwhile, the longest current numerical simulations start with $G\,
\Mtot\, \omegah / c^{3} \gtrsim \num{0.035}$.  The upper-left panel of
Fig.~\ref{fig:MismatchContours} shows that hybrids created using such
simulations would only be sufficient for $\Mtot \gtrsim
\SI{26}{\SolarMass}$.  Of course, real detector data is searched for a
range of system parameters; a real $\SI{10}{\SolarMass} +
\SI{10}{\SolarMass}$ system might be detected by an inaccurate
$\SI{6}{\SolarMass} + \SI{18}{\SolarMass}$ template, for
example~\cite{BoyleEtAl:2009}.\footnote{In much of the literature,
  this difference is highlighted by distinguishing between
  ``effectualness'' (the match between a given signal and the best fit
  in a template bank) and ``faithfulness'' (the match between a given
  signal and the particular signal in the template bank with the same
  physical parameters).}  %
This dramatically reduces the accuracy requirements on a template bank
for detection~\cite{BuonannoEtAl:2009}.  However, while template banks
may be subject to loose accuracy requirements, numerical relativity
will generally be used for other purposes---most likely calibrating
template banks to NR waveforms or hybrids incorporating them.  The
results presented above show that any such calibration is bound to
exhibit very large errors for low-mass systems unless the numerical
simulation is very long.

More stringent demands are placed on waveforms for parameter
estimation, depending on the SNR of the observed signal.  For example,
modeling the unequal-mass high-spin system to high accuracy would
require simulating nearly the entire in-band signal; the simulation
would need to begin roughly \SI{40}{\percent} above the seismic wall
to achieve mismatches of $\Mismatch \approx \num{e-4}$.  These grim
results present discouraging prospects for accurate modeling of
precessing systems, PN approximations for which are known to
still-lower order.

On the other hand, for small values of the mismatch, the appropriate
value of $\omegah$ varies almost linearly with $\Mtot$.  The initial
orbital frequency $\Omega_{0}$ required for a simulation will then be
nearly proportional to the total mass of the modeled system, so the
length of the simulation will vary roughly as $\Mtot^{-8/3}$.  This
strong dependence shows clearly that target accuracies for a
simulation to be used across a range of masses should include
carefully considered mass dependence.  Such dependence is incorporated
into the technique for determining $\Omega_{0}$ in
Sec.~\ref{sec:TargetAccuracyAsFunctionOfMass}.  For the $q=\num{10}$
systems in Fig.~\ref{fig:MismatchContours}, this improves the
situation dramatically.  The smallest black hole observed to date has
$M \gtrsim \SI{3}{\SolarMass}$~\cite{GelinoHarrison:2003}, so a binary
with $q=\num{10}$ would have $\Mtot \gtrsim \SI{33}{\SolarMass}$,
substantially raising the value of $\omegah$ required to achieve a
given mismatch for astrophysically likely sources.  In this sense,
systems with large mass ratios are actually easier to model than
comparable-mass systems.

There are several possible flaws in these uncertainty estimates.  Most
basically, we simply assume that the uncertainty in our range of
waveforms is a suitable proxy for the error in the waveforms.  Of
course, these models---both PN and EOB---may simply be wrong.  For
example, we can imagine that some fundamental error exists in our
understanding of approximations to Einstein's equations for black-hole
binaries.  In that case, our models may be perfectly precise but
entirely inaccurate; the exact waveform would lie outside the bounds
of our uncertainty estimate.  Moreover, these estimates depend on the
assumption that the range of plausible hybrid waveforms is neither too
narrow nor too broad.  The choices made above were based largely on
coincidences of history; some other reasonable, equally accurate, but
not-yet-imagined approximant may exist, lying far from the
approximants used here.  Conversely, there may be some subtle error in
one or more members of this group of approximants, leading to
unnecessarily large uncertainty.  Unfortunately, the only obvious way
to detect such errors is to test the results using very long and
accurate numerical simulations.

Taken together, these results indicate that more work will be needed
to produce accurate waveforms for stellar-mass black-hole binaries,
even for aligned-spin systems.  Improvements may come in the form of
higher-accuracy PN or EOB waveforms, longer numerical simulations, or
both.  This paper has not treated precessing systems simply because
the production of full waveforms for such systems is still in its
infancy.  No doubt, however, the uncertainties are greater than in the
cases discussed above.  While both analytical and numerical relativity
have clearly made great progress in the past decade, much remains to
be done.

\begin{acknowledgments}
  It is my pleasure to thank Ilana MacDonald and Larne Pekowsky for
  useful discussions and direct comparisons of the results of our
  various codes; Larry Kidder, Harald Pfeiffer, B{\'{e}}la
  Szil{\'{a}}gyi, and Saul Teukolsky for very useful discussions;
  Alessandra Buonanno and Yi Pan for kindly sharing their knowledge of
  EOB models; Larne Pekowsky and Duncan Brown for providing helpful
  information on noise curves and accuracy requirements; P. Ajith,
  Sascha Husa, and the rest of the \mbox{NINJA-2} collaboration for
  help validating the Taylor PN approximants used in this work.  This
  project was supported in part by a grant from the Sherman Fairchild
  Foundation; by NSF Grants No.\ PHY-0969111 and No.\ PHY-1005426; and
  by NASA Grant No.\ NNX09AF96G. The numerical computations presented
  in this paper were performed primarily on the \texttt{Zwicky}
  cluster hosted at Caltech by the Center for Advanced Computing
  Research, which was funded by the Sherman Fairchild Foundation and
  the NSF MRI-R\textsuperscript{2} program.
\end{acknowledgments}

\appendix 

\section{The (un)importance of the choice of ersatz NR waveform}
\label{sec:ImportanceOfErsatzNRWaveform}
Section~\ref{sec:Motivation} presents arguments that the mismatch
between two hybrid waveforms with ersatz NR data should be almost
completely insensitive to the phasing of the ersatz waveform above the
hybridization frequency, and only weakly dependent on the amplitude.
The key point is that the two hybrids are identical for $\omega >
\omegah$.  It was thus argued that the particular choice of ersatz NR
waveform should not strongly affect the results, as long as the power
is fairly correctly distributed in the frequency domain.

One simple way to check this claim is to use a different ersatz NR
waveform.  Here, we will reproduce the crucial result of
Fig.~\ref{fig:MismatchContours} in the equal-mass nonspinning case
with different ersatz NR data.  In this case, we will substitute the
EOB waveform with a numerical waveform hybridized with TaylorT4 to
extend to lower frequencies.  The numerical waveform is the same one
introduced in Ref.~\cite{ScheelEtAl:2009}, except that
Regge--Wheeler--Zerilli wave extraction is used to produce $h$.  The
waveform is hybridized exactly as in Ref.~\cite{BoyleEtAl:2009}.  This
hybrid is then substituted for the EOB waveform wherever it is called
for.

\begin{figure}
  \includegraphics[width=\linewidth]{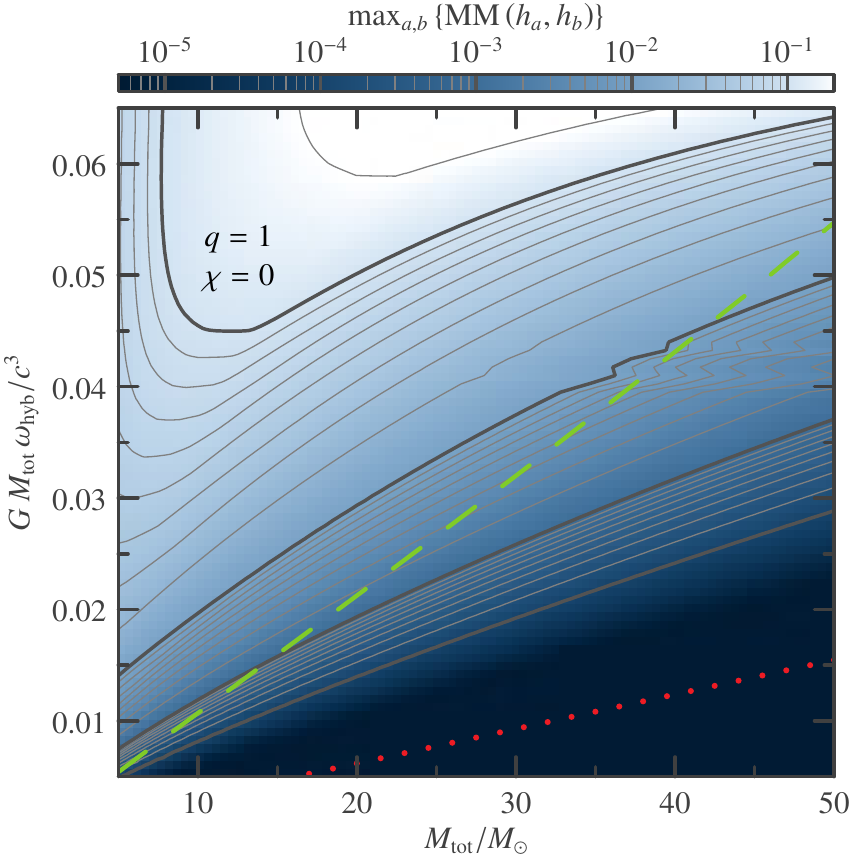}
  \caption{\CapName{Using real NR data for merger and ringdown} This
    plot reproduces the top-left plot of
    Fig.~\ref{fig:MismatchContours}, using real NR data in place of
    the EOB ersatz NR waveform.  The NR data starts at frequency
    $\omega \approx 0.035$, and is extended to lower frequencies by
    hybridizing with a TaylorT4 waveform.  The plots are almost
    identical, indicating that the procedure is not strongly sensitive
    to details of the ersatz NR waveform.  \label{fig:q1_chis0_NR}}
\end{figure}

The results are shown in Fig.~\ref{fig:q1_chis0_NR}.  Comparing with
the upper-left plot of Fig.~\ref{fig:MismatchContours}, we see
excellent agreement throughout the plot.  The plot shown here does
exhibit some jagged lines in the range $\num{0.04} < G\, \Mtot\,
\omegah / c^{3} < \num{0.045}$.  These are evidently due to noise in
the waveform itself, which appears to be related to junk radiation.
That noise can easily lead to imperfect hybrids, especially using the
frequency-alignment scheme of
Eq.~\eqref{eq:PhaseAndFrequencyMatching}.

\begin{figure}
  \includegraphics[width=\linewidth]{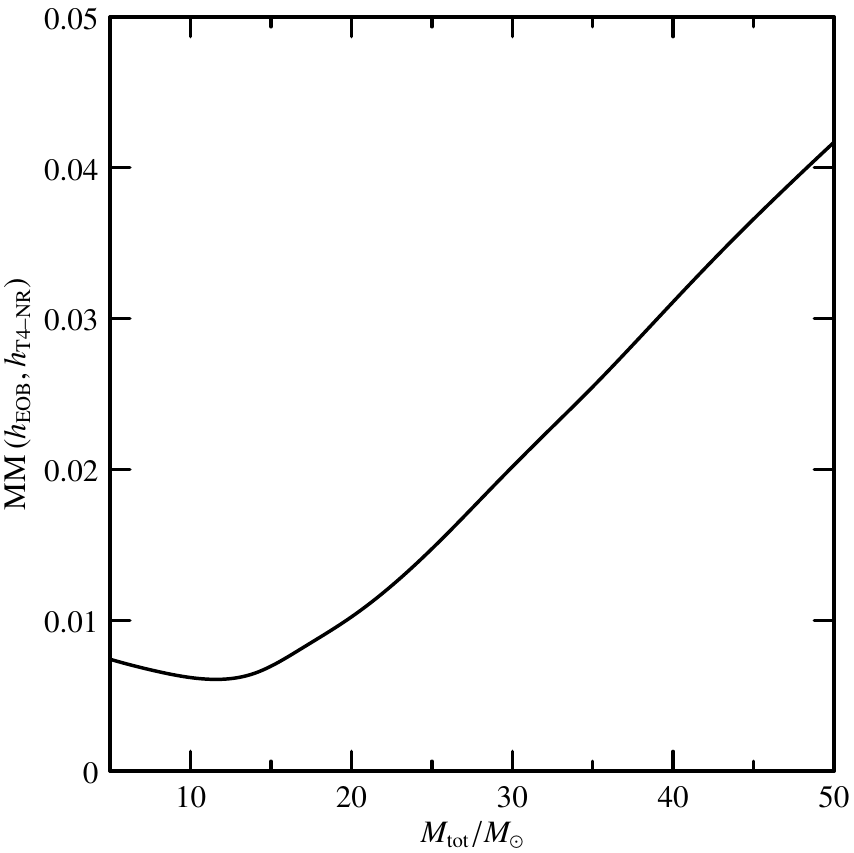}
  \caption{\CapName{Mismatch between the EOB waveform and the NR
      hybrid} This plot shows the mismatch as a function of total mass
    between the EOB waveform used in the body of this paper and the NR
    hybrid used in Fig.~\ref{fig:q1_chis0_NR}.  The similarity between
    Figs.~\ref{fig:MismatchContours} and~\ref{fig:q1_chis0_NR} despite
    the significant mismatches shown here lead us to conclude that the
    uncertainties shown in those figures are indeed robust with
    respect to the choice of ersatz NR waveform.  \label{fig:EOBvsPN}}
\end{figure}

At the very least, this demonstrates that the simplistic
ringdown-alignment technique used for the EOB waveform in this paper
(see Sec.~\ref{sec:EOBModel}) does not significantly affect the final
results.  On the other hand, we might worry that the NR hybrid used
here is practically identical to the EOB waveform used in the main
text of this paper, because the EOB waveform aligns quite accurately
to the very late stages of the NR data.  In fact, that alignment is
misleading, because it requires coherence over the relatively short
span of the numerical data.  Judged in terms of the mismatch, the NR
hybrid and the EOB waveform are quite distinct, shown in
Fig.~\ref{fig:EOBvsPN} as a function of the total mass of the system.

\section{Details of the implementation}
\label{sec:DetailsOfTheImplementation}
The results of this paper depend sensitively on accurate numerical
implementation of the technique.  The various approximants, their
hybrids, and the mismatch must all be calculated to high accuracy to
ensure that the plots of Fig.~\ref{fig:MismatchContours} depict
uncertainty in the models, rather than errors in the numerical
methods.  This section outlines the steps necessary to obtain accurate
results.  In short, every attempt was made to ensure that the model
waveforms were as accurate as possible, and each number quoted in this
section was tested to ensure that making it more stringent had no
significant effect on the final results.

\subsection{Post-Newtonian ingredients}
\label{sec:PNModel}
The TaylorT$n$ approximants used in this paper are based on the
results of Ref.~\cite{NINJA2:2011}.  In the appendix of that
reference, the most current and complete PN results are collected and
expressed in consistent notation.  In particular, the orbital energy,
tidal heating, and gravitational-wave flux are given to \num{3.5}-PN
order in nonspinning terms, and incomplete \num{2.5}-PN order in
spinning terms.  The spins are assumed to be aligned or anti-aligned
with the orbital angular momentum.  These are the basic ingredients to
construct the phasing of TaylorT1--T4 approximants, as described
succinctly in Ref.~\cite{BoyleEtAl:2007}.  The orbital phase thus
derived is then used in the waveform amplitudes described by
Ref.~\cite{NINJA2:2011}, which include nonspinning terms up to
\num{3}-PN order, and spinning terms through \num{2}-PN order.

The main practical concern in constructing these waveforms is
producing data on a sufficiently fine grid that accurate derivatives
are available for the alignment step,
Eq.~\eqref{eq:PhaseAndFrequencyMatching} of the hybridization
procedure.  For the TaylorT1 and T4 models, this is accomplished by
setting a tight tolerance on the numerical integration scheme, as
discussed in Sec.~\ref{sec:NumericalIntegrationOfODES}.  For TaylorT2,
the waveform is evaluated on at least \num{50000} uniformly
distributed values of the velocity parameter $v$, which is the
independent variable of this model, ranging up to $v=\num{1}$.
Similarly, the TaylorT3 waveform was evaluated at \num{50000} values
of the independent time parameter $\tau$, distributed at roughly
uniform intervals of $v$.  Lower values of these numbers led to poor
hybrids, characterized by noise at low values of $\omegah$ in the
plots of Fig.~\ref{fig:MismatchContours}.

\subsection{EOB model}
\label{sec:EOBModel}
The EOB model used for this paper was designed to incorporate recent
improvements to the inspiral portion of the model, including spin
terms, while also remaining robust, allowing its application to the
somewhat extreme case of $q=\num{10}$, $\chi=\num{0.95}$.  The primary
compromises made in the interests of robustness were abandonment of
the factorized multipolar waveforms of Ref.~\cite{PanEtAl:2010b} and
coherent attachment of the ringdown portion of the waveform.  The
former compromise requires the use of the Pad{\'{e}}-expanded flux to
calculate the phasing of the system, and the standard PN multipolar
waveforms.  These are both reasonable substitutions: the flux term is
used in the EOB code for the LIGO Algorithm Library; the PN multipolar
waveform should still be accurate for most of the
inspiral~\cite{PanEtAl:2010b}.  The latter compromise primarily
affects the phase of the waveform during its very last stages.  As was
argued in Sec.~\ref{sec:Motivation}, this is unlikely to have any
significant effect.  In any case, the uncertainty of the plausible
waveforms is dominated by the TaylorT$n$ approximants in all cases
shown in this paper, and essentially identical results are obtained
when using real numerical data for the merger and ringdown, suggesting
that any error in the EOB model does not affect the results.

The EOB Hamiltonian used here is roughly the same as the one given by
Ref.~\cite{PanEtAl:2010}, except that nonspinning terms in the metric
functions $A(r)$ and $D(r)$ are extended with new terms from
Ref.~\cite{PanEtAl:2011}.  Thus, in the nonspinning case, the
Hamiltonian of this EOB model reduces exactly to the Hamiltonian of
Ref.~\cite{PanEtAl:2011}; in the spinning case, it reduces nearly to
the Hamiltonian of Ref.~\cite{PanEtAl:2010}.  The angular momentum
flux is described by Eq.~(65) of Ref.~\cite{BoyleEtAl:2008}, where the
term $F^{4}_{4}$ is given by the \Pade expansion of the flux from
Ref.~\cite{NINJA2:2011}.  The standard formula for $v_{\text{pole}}$
gives very poor results for high spins.  For this paper, the following
extension of $v_{\text{pole}}$ to the spinning case is used:
\begin{equation}
  \label{eq:vPole}
  v_{\text{pole}} = \frac{6 + 2\, \nu} {\sqrt{(3 + \nu)\,(36 - 35\,
      \nu)} - \chi_{s}\, (8 - 4\, \nu)}~.
\end{equation}

Initial data is set according to Eqs.~(4.6) and~(4.13) of
Ref.~\cite{BuonannoDamour:2000}.  Eccentricity is then iteratively
reduced to $e \leq 10^{-14}$ using Eqs.~(71) and~(73) of
Ref.~\cite{BuonannoEtAl:2010}.  For the high-spin cases, this method
does not work directly.  Instead, the spin is increased in stages.
The non-eccentric initial data for the given mass ratio is first
obtained with $\chi=0$, then used as initial data for eccentricity
reduction with $\chi=\num{0.1}$.  This is repeated, incrementing the
value of $\chi$, until the desired spin parameter is reached.  That
non-eccentric initial data is then used to evolve the full inspiral.
Reducing eccentricity is not only more faithful to the scenario
modeled by the other approximants, but also allows larger time steps
to be taken by the numerical integration scheme; significant
eccentricity would require at least a few steps to be taken per orbit.

The integration ends when the EOB radial parameter is smaller than
\num{1}, or the radial momentum becomes positive.  In all cases
explored for this paper, the amplitude of the resulting waveform
reaches a peak, roughly where merger is expected, and roughly similar
in amplitude to the peak expected from numerical simulations.
Previously published EOB models align a sum of decaying quasinormal
modes to the rising side of this peak~\cite{BuonannoDamour:2000,
  BuonannoEtAl:2010, PanEtAl:2011}.  Those techniques do not seem to
be sufficiently robust to apply naively to the extreme cases discussed
in this paper.  Moreover, such techniques seem to be unnecessary; as
argued previously, the particular details of the end of the waveform
will not strongly affect the final results, especially for the small
portion of the waveform represented by ringdown.  For these reasons, a
simple---though undoubtedly inaccurate---method is used to attach a
single quasinormal mode to the inspiral waveform.  The descending side
of the amplitude peak is used, and the quasinormal mode with the
longest decay time is attached at the unique point such that the
amplitude and its first derivative are continuous.

\subsection{Numerical integration of ODEs}
\label{sec:NumericalIntegrationOfODES}
The TaylorT1, TaylorT4, and EOB waveforms are integrated numerically
by the eighth-order Dormand--Prince method implemented in
\textit{Numerical Recipes}~\cite{PressEtAl:2007}.  In all cases, the
absolute tolerance was set to $\texttt{atol} = 0$ because of the
vastly different scales of the dependent variables.  The value of the
relative tolerance $\texttt{rtol}$ was chosen by looking at the
convergence of the phase of each approximant.  Tolerances from
$\num{e-4}$ to $\num{e-11}$ gave the same results to within small
fractions of a radian over the entire \SI{\sim 100000}{\radian}
inspiral.  For EOB, $\texttt{rtol} = \num{e-6}$ was chosen to be
conservative, while still allowing the code to run very quickly (less
than one second per waveform).  For TaylorT1 and T4, $\texttt{rtol} =
\num{e-10}$ was used as a crude but effective way of ensuring that
output was frequent enough to produce smooth derivatives for the
alignment procedure, Eq.~\eqref{eq:PhaseAndFrequencyMatching}.
Additionally, dense output was used to save \num{50} intermediate
points per time step, which further improved the alignment procedure.
Practically identical results were obtained with the Bulirsch--Stoer
integration scheme, except that this method could not reliably
continue into the delicate final few radians of the EOB integration.
Integration continues until the dependent variables or their
derivatives reach some unphysical value: for TaylorT1 and T4, angular
frequency is required to remain positive; for EOB, radial momentum is
required to remain negative, and radius greater than \num{1}.

\subsection{Fourier transforms and mismatches}
\label{sec:FourierTransforms}
Fourier transforms find two applications in the calculation of the
mismatch.  First, and most obviously, time-domain waveforms must be
converted to the frequency domain for use in the inner product,
Eq.~\eqref{eq:InnerProduct}.  Second, the match itself is then
evaluated by taking an inverse Fourier transform.  Assuming that the
waveforms $h_{a}$ and $h_{b}$ are normalized so that
$\InnerProduct{h_{a} | h_{a}} = \InnerProduct{h_{b} | h_{b}} = 1$, and
combining expression~\eqref{eq:PhaseDifferenceChange} with
Eq.~\eqref{eq:Overlap}, we see that
\begin{subequations}
  \begin{align}
    \label{eq:OverlapByFFT}
    \Overlap{h_{a} | h_{b}} %
    &= \max_{\Delta T, \Delta \Phi}\, \InnerProduct{h_{a} | h_{b}}
    \\ %
    &= 2\, \max_{\Delta T, \Delta \Phi}\, \Re\int_{-\infty}^{\infty}\,
    \frac{\tilde{h}_{a}\, \tilde{h}_{b}^{\ast}} {S_{n}(\abs{f})}\,
    e^{2\, i\, \sgn(f)\, \Delta\Phi + 2\, \pi\, i\, f\, \Delta T}\, df
    \\ %
    &= 4\, \max_{\Delta T}\, \abs{\int_{0}^{\infty}\,
      \frac{\tilde{h}_{a}\, \tilde{h}_{b}^{\ast}} {S_{n}(\abs{f})}\,
      e^{2\, \pi\, i\, f\, \Delta T}\, df }
  \end{align}
\end{subequations}
Note that the integral in the last expression is simply the inverse
Fourier transform of $\tilde{h}_{a}\, \tilde{h}_{b}^{\ast} /
S_{n}(\abs{f})$.  The maximization over $\Delta T$ involves selecting
the largest element (in absolute value) of the discrete set produced
by the fast Fourier transform of that quantity.

Two concerns drive the application of these Fourier transforms:
aliasing at high frequencies, and Gibbs artifacts at low frequencies.
To avoid aliasing, the sampling interval of the time-domain waveforms
must be set by the highest frequency of the lowest-mass system of
interest.  In our case, that system has a mass of $\Mtot =
\SI{5}{\SolarMass}$.  An acceptable sampling frequency is half the
Advanced LIGO sampling frequency: $f_{\text{s}} = \SI{8192}{\hertz}$.
On the other hand, avoiding Gibbs artifacts at low frequencies
requires the waveforms to start early enough that the waveform ``turns
on'' outside of the LIGO band, and its amplitude is very small at that
point.  Tests with the waveforms used in this paper show that an
initial frequency of \SI{8}{\hertz} is sufficient to ensure accuracy
of the mismatch to $\Mismatch \lesssim 10^{-7}$.  For $\Mtot =
\SI{5}{\SolarMass}$, this corresponds to a dimensionless initial
orbital frequency of $G\, \Mtot\, \Omega_{0} / c^{3} \approx
\num{1.97e-5}$.  The waveforms used in this paper were calculated in
dimensionless units, starting with that frequency, hybridized as
necessary, scaled to the appropriate total mass, projected to the
positive $z$ axis, and interpolated to a uniform time grid with
spacing $\Delta t = 1/f_{\text{s}}$.  In extreme cases, these
waveforms can consume hundreds of megabytes each.  Given that five
such waveforms need to be compared, and that comparison requires
significant additional memory, the full memory usage can easily reach
several gigabytes.  Because of the large memory requirements, the
calculations for Fig.~\ref{fig:MismatchContours} were performed on a
cluster having ample memory in a single node.  To use CPU resources
efficiently OpenMP~\cite{DagumAndMenon:1998} was employed, which
allows very simple alterations of source code to incorporate multiple
processes---just three additional lines of code enabled
multiprocessing which resulted in a speed improvement by a factor of
four.



\vspace{.1in}
\vfil
\let\c\Originalcdefinition
\let\d\Originalddefinition
\let\i\Originalidefinition
\bibliography{Externals/References}


\end{document}